\documentclass[12pt,preprint]{aastex}

\usepackage{graphicx, color}% Include figure files
\usepackage{dcolumn}% Align table columns on decimal point
\usepackage{bm}% bold math
\usepackage{booktabs}
\usepackage{natbib}
\usepackage{lscape}

\newcommand {\inty}[2]{\int_{#1}^{#2}}

\newcommand {\pdif}[3][]{\frac{\partial^{#1}#2}{\partial#3^{#1}}}
\newcommand {\abs}[1]{\left|#1\right|}
\newcommand {\fr}{\frac}
\newcommand {\lsim}{\hspace{0.3em}\raisebox{0.4ex}{$<$}\hspace{-0.75em}\raisebox{-.7ex}{$\sim$}\hspace{0.3em}}
\newcommand {\gsim}{\hspace{0.3em}\raisebox{0.4ex}{$>$}\hspace{-0.75em}\raisebox{-.7ex}{$\sim$}\hspace{0.3em}}

\newcommand {\mart}[2][]{\sqrt[#1]{\mathstrut{#2}}}
\newcommand{\myemail}{takashim@eps.s.u-tokyo.ac.jp}
\newcommand{\Yohkoh}{\it Yohkoh}
\newcommand{\FP}{Fokker-Planck}
\newcommand{\TPP}{trap-plus-precipitation}
\newcommand{\bremss}{bremsstrahlung}
\newcommand{\gyros}{gyrosynchrotron}
\newcommand \gammat[2] {\gamma_{\rm #1}^{\rm #2}(t)}
\newcommand \deltat[1] {\delta_{\rm #1}(t)}
\newcommand \Deltat[1] {\Delta_{\rm #1}(t)}
\newcommand \alphat[1] {\alpha^{\rm #1}(t)}

\begin{document}

\shorttitle{Non-thermal Emissions and Electrons in the Flare}
\shortauthors{Minoshima et al.}

\title{Comparative Analysis of Non-thermal Emissions and Study of Electron Transport in a Solar Flare}

\author{T. Minoshima\altaffilmark{1}, T. Yokoyama\altaffilmark{1}, and N. Mitani\altaffilmark{2}}
% \affil{
% Department of Earth and Planetary Science, University of Tokyo,
% 7-3-1, Hongo, Bunkyo, Tokyo, 113-0033 Japan;
% }
\altaffiltext{1}{
Department of Earth and Planetary Science, Graduate School of Science, University of Tokyo,
7-3-1, Hongo, Bunkyo-ku, Tokyo, 113-0033 Japan;
}
\altaffiltext{2}{
Institute of Astronomy, University of Tokyo, 
2-21-1, Osawa, Mitaka, Tokyo, 181-0015, Japan;
}
\email{\myemail}
%\affil{
%Department of Earth and Planetary Science, University of Tokyo,
%7-3-1, Hongo, Bunkyo, Tokyo, 113-0033 Japan;
%yokoyama.t@eps.s.u-tokyo.ac.jp
%}

%\date{\today}

\begin{abstract}
% Observations of the non-thermal emissions, hard X-rays (HXRs) and optically thin microwaves, have direct information on the accelerated (non-thermal) electrons in solar flares. Because effective energies of the HXR and microwave emitting electrons are different, the observations using both wavelengths are powerful mean for discussing physics of flare non-thermal electrons over a wide range of electron energies.
 We study the non-thermal emissions in a solar flare occurring on 2003 May 29 by using RHESSI hard X-ray (HXR) and Nobeyama microwave observations. This flare shows several typical behaviors of the HXR and microwave emissions: time delay of microwave peaks relative to HXR peaks, loop-top microwave and footpoint HXR sources, and a harder electron energy distribution inferred from the microwave spectrum than from the HXR spectrum. In addition, we found that the time profile of the spectral index of the higher-energy ($\gsim 100$ keV) HXRs is similar to that of the microwaves, and is delayed from that of the lower-energy ($\lsim 100$ keV) HXRs. We interpret these observations in terms of an electron transport model called {\TPP}. We numerically solved the spatially-homogeneous {\FP} equation to determine electron evolution in energy and pitch-angle space. By comparing the behaviors of the HXR and microwave emissions predicted by the model with the observations, we discuss the pitch-angle distribution of the electrons injected into the flare site. We found that the observed spectral variations can qualitatively be explained if the injected electrons have a pitch-angle distribution concentrated perpendicular to the magnetic field lines rather than isotropic distribution. %  Our calculations suggest that the injected electrons which pitch-angle distribution is elongated transverse to magnetic field lines rather than be isotropic are somewhat favorable to qualitatively explain the observations, but quantitatively insufficient.
\end{abstract}


\section{Introduction}\label{sec1}
 Observations of hard X-rays (HXRs), microwaves, and occasionally gamma-rays in solar flares tell us that a significant amount of non-thermal particles are produced. Among them, HXR and microwave observations are believed to provide the most direct information on electrons. Because HXRs below $\sim 100$ keV are emitted primarily by electrons with energy below several hundred keV via {\bremss} radiation \citep{1971SoPh...18..489B}, whereas microwaves above $\sim 10$ GHz are emitted by electrons above several hundred keV via {\gyros} \citep{1969ApJ...158..753R,1999spro.proc..211B}, these two sources of emission give us information on electrons in two different energy ranges. Therefore, a comparative study by using both HXR and microwave observations is useful for discussing the physics of flare non-thermal electrons over a wide range of energies.

 Impulsive behavior is commonly seen in both HXR and microwave lightcurves \citep{1974IAUS...57..105K}, but the two emissions do not necessarily behave identically. Temporally, higher-energy HXR and microwave emissions tend to be delayed from lower-energy HXRs \citep[e.g.,][]{1978ApJ...223..620C,1983Natur.305..292N,1985ApJ...292..699B,1997ApJ...487..936A}. \cite{1997ApJ...487..936A} statistically analyzed the low-pass filtered HXR lightcurves for 78 flares observed with the {\it Compton Gamma Ray Observatory} ({\it CGRO}) and find a systematic increase of time delay toward higher energy. They interpreted these time delays in terms of electron precipitation under Coulomb collisions.

Spatially, microwave sources do not always coincide with HXR sources. HXRs are typically emitted at the footpoint regions of the flare loop \citep{1994PhDT.......335S} whereas microwaves are emitted mainly at the loop-top region \citep{2002ApJ...580L.185M}. \cite{2002ApJ...580L.185M} suggested that only electrons with a pancake pitch-angle distribution concentrated transverse to the magnetic field lines can explain the observed loop-top microwave source.
% discussed the pitch-angle distribution of the electrons injected at the flare loop based on their microwave observations. They 
%  suggested that only the injected electrons which pitch-angle distribution is elongated transverse to magnetic field lines (pancake pitch-angle distribution) can meet the well-observed loop-top microwave source.

 Spectrally, \cite{2000ApJ...545.1116S} statistically studied the correlation of the HXR and microwave spectral indices for 57 peaks of the non-thermal emission in 27 flares. They found that the electron energy distribution inferred from the microwave spectrum is systematically harder than that inferred from the HXR spectrum, and suggested that the electron energy distribution becomes harder towards higher energy. There are three probable explanations for such spectra: (1) two (or more) different electron populations with distinct physical characteristics, (2) ``second-step acceleration'' \citep[e.g.,][]{1976SoPh...49..343B}, and (3) ``{\TPP} (TPP)'' \citep[e.g.,][]{1976MNRAS.176...15M}. 

\cite{1976MNRAS.176...15M} presented analytic solutions of the electron energy continuity equation under two conditions: {\it strong} and {\it weak diffusion limits} \citep{1966JGR....71....1K}. In the strong diffusion limit, electrons injected into the loop undergo significant scattering and then are quickly isotropized during the loop transit. They can escape from the loop with a precipitation rate proportional to their velocity, $\nu_{\rm p} \propto v$. In the weak diffusion limit, on the other hand, electrons are less scattered during the transit. When the loss cone distribution is formed, the pitch-angle diffusion time $\tau_{\rm d}$, which is longer than the transit time, controls the electron precipitation, yielding $\nu_{\rm p} \propto 1/\tau_{\rm d}$. The precipitation rate and the evolution of electrons vary, depending on which condition applies.

% The TPP model may explain the discrepancy of the energy distribution between the HXR and microwave emitting electrons reported by \cite{1997SoPh..175..157S,2000ApJ...545.1116S}. Their interpretation of the observations is subject to the assumption that the microwave sources spatially coincide with the HXRs or that the strong diffusion limit is adequate. However, as reported by \cite{2002ApJ...580L.185M}, HXR and microwave sources do not necessarily coincide with each other. In this case, the HXR emitting electrons may have a different energy distribution from the microwave emitting ones. Imaging as well as spectral data of the non-thermal emissions are required to confirm the role of TPP on the parent electrons.

There have been many observations that can be explained in terms of the TPP model \citep[e.g.,][]{2000ApJ...531.1109L}. \cite{2002ApJ...576L..87Y} reported the Nobeyama Radioheliograph (NoRH; \citealt{1994PROCIEEE...82..705}) observation of a flare occurring on 1999 August 28. The NoRH observation showed clear flare-loop structure and propagating features along the loop. They showed that the microwave spectrum in the optically-thin regime is hard (with spectral index $\sim 1.5$) around the loop-top, and then becomes softer (spectral index $\sim 3.5$) toward the footpoints. Their observation indicates that the higher-energy electrons are efficiently trapped within the loop, supporting the TPP model. For this event, however, no HXR observation was available for comparison with the microwave observation. \cite{2000ApJ...545.1116S} pointed out that the discrepancy of the energy distribution between the HXR and microwave emitting electrons found in their study could be explained by the TPP model. In their study, there was no imaging observation to confirm their suggestion. If the HXR and microwave sources do not coincide spatially, the discrepancy of the energy distribution between the HXR and microwave emitting electrons can be explained by the different spatial distribution between the HXR and microwave emitting electrons as a result of the TPP model. Imaging as well as spectral data at both HXR and microwave wavelengths are essential to confirm the role of TPP on the parent electrons

%  As mentioned above, the comparative study of flare non-thermal emissions is a powerful mean to extract the information on electrons and to discuss their acceleration and transport mechanism over a wide range of energies.
In this paper we analyze the non-thermal emissions of the 2003 May 29 flare by using the {\it Reuven Ramaty High Energy Solar Spectroscopic Imager} \citep[RHESSI;][]{2002SoPh..210....3L}, the Nobeyama Radio Polarimeters \citep[NoRP,][and references theirin]{1985PASJ...37..163N} and NoRH. RHESSI has superior spectroscopic ability from $\sim 3$ keV to $\sim 17$ MeV, providing the HXR spectrum from $\sim 3$ keV to $\sim 300$ keV with a spectral resolution of $\sim 1$ keV and arbitrary energy bands.
In previous studies, the temporal evolution of the (HXR) spectrum has been considered in less detail, probably due to instrumental limitations. However, the temporally-resolved analysis of the spectrum is important because non-thermal emissions and thus non-thermal electrons are the most ``time-varying'' objects in solar flares. RHESSI enables us to analyze an accurate, temporally-resolved HXR spectrum above $\sim 100$ keV.
 Because the HXRs above $\sim 100$ keV are mainly emitted by electrons above $\sim 200$ keV \citep{1996ApJ...464..974A}, RHESSI's well-resolved spectral data below $\sim 300$ keV provides us more accurate information on electrons from tens to hundreds of keV than before. Combining the RHESSI HXR and NoRH/NoRP microwave spectral data allows us to fully cover the electrons from tens to thousands of keV.

 For a physical interpretation of the observations, we use a numerical model of TPP which treats the pitch-angle diffusion more generally than the analytic solutions developed for the weak and strong diffusion limits. \cite{2000ApJ...543..457L} performed a similar treatment of the electron transport to explain their microwave observation of a flare on 1993 June 3. We also predict the microwave and HXR emissions from the calculated electron distribution. Comparing these model results with the observations, we discuss electron injection and transport, and address how the pitch-angle distribution of the injected electrons affects the evolution of the trapped and precipitating electrons, and their resultant emissions. 

The paper proceeds as follows. In {\S}~\ref{sec2} we present a comparative study of the non-thermal emissions of a solar flare occurring on 2003 May 29, by using the RHESSI HXR and Nobeyama microwave observations. Temporally-resolved spectra of the HXRs and microwaves are analyzed in detail. We discuss energy-dependent delays of the time profiles of the spectral indices, which have not been discussed in previous studies. In {\S}~\ref{sec3} we present our treatment of the TPP model. We numerically solve the spatially-homogeneous {\FP} equation \citep{1990A&A...230..213M} with the Coulomb interaction \citep[e.g.,][]{1981ApJ...251..781L} and a time-dependent injection. In {\S}~\ref{sec4} we describe the time evolution of the trapped and precipitating electron distribution and the predicted microwave and HXR emissions. The behavior of the HXR and microwave emissions predicted by the model are compared with the observations, allowing us to give some constraints on the properties of flare non-thermal electrons. In {\S}~\ref{sec5} we conclude our study. %In {\S}~\ref{sec4} we summarize our study and discuss a probable physics of flare non-thermal electrons based on both the observations and the calculations.

\section{Observations}\label{sec2}
We studied a solar flare that occurred on 2003 May 29. The GOES soft X-ray (SXR) level was X1.2 (Fig. \ref{ltc}; {\it upper}). This flare occurred at S07W31 at 00:50 UT, and lasted about 1 hour. RHESSI detected a significant amount of high energy ($< 300$ keV) HXRs during 01:00 - 01:05 UT; we define this period as the impulsive phase. NoRP and NoRH also observed this flare.

\subsection{Lightcurves}\label{sec2_1}
\begin{figure}[htbp]
\centering
\epsscale{0.75}
\plotone{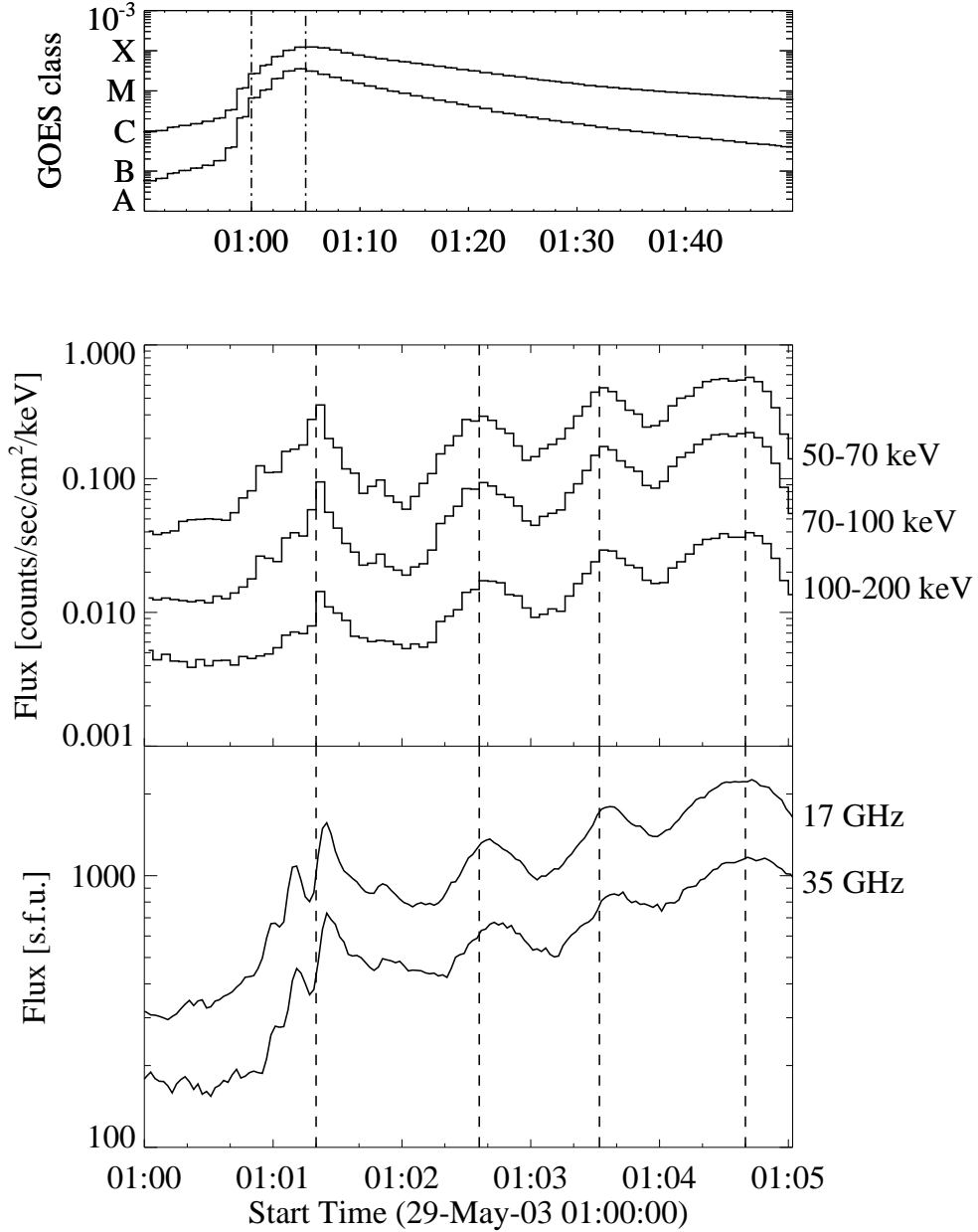}
\caption{Lightcurves for the 2003 May 29 flare. {\it Upper}: The GOES SXR lightcurves during 00:50 - 01:50 UT. The period during 01:00 - 01:05 UT (dot-dashed lines) is defined as the impulsive phase. {\it Lower}: Lightcurves of the non-thermal emissions during the impulsive phase. The upper part shows the HXR lightcurves (in units of counts/${\rm cm^2}$/sec/keV) taken with RHESSI in three energy bands: 50-70 keV, 70-100 keV, and 100-200 keV from top to bottom. The lower part shows the microwave lightcurves (S.F.U.) observed in the NoRP 17 GHz and 35 GHz bands from top to bottom. The dashed lines denote the peak times of each spike in the 50-70 keV band.}
\label{ltc}
\end{figure}

% Fig. \ref{ltc} shows the lightcurves of the non-thermal emissions during 01:00 - 01:05 UT. The upper panel shows the HXR lightcurves (in units of counts/${\rm cm^2}$/sec/keV) taken with RHESSI in three energy bands, 40-60 keV, 60-80 keV, and 100-150 keV from top to bottom. The lower panel shows the microwave lightcurves (in S.F.U.) observed with the NoRP 17 GHz and 35 GHz from top to bottom. The dashed lines denote the peak times of each spike in the RHESSI 40-60 keV. The lightcurves consist of four spikes which durations are $\sim 1$ minute. The spiky behavior seems to be common between the HXRs and microwaves. This indicates that the 17 GHz and 35 Ghz microwaves are the non-thermal {\gyros} emissions. But the microwaves show slight delays from the HXRs. % So we simply check the lightcurve delays by comparing the peak times of each spike in two energy bands (frequencies). 
% The peak times of the HXR lightcurves within 50-200 keV almost coincide within their temporal resolution of 2 seconds. But the microwave lightcurves delay by about 4 seconds (on average) from the HXRs.
Figure \ref{ltc} shows the lightcurves of SXRs ({\it upper}), HXRs and microwaves ({\it lower}) during the impulsive phase (01:00 - 01:05 UT). The upper part of the lower plot shows the HXR lightcurves taken with RHESSI in three energy bands: 50-70 keV, 70-100 keV, and 100-200 keV. The dashed lines denote the peak times of each spike in the 50-70 keV band. The HXR lightcurves consist of four spikes, each with durations of $\sim 1$ minute. The peaks of the HXR lightcurves in the three energy bands coincide with each other within a temporal resolution of 2 s.

The microwave lightcurves at 17 GHz and 35 GHz presented in the lower part of the lower plot in Figure \ref{ltc} also show four peaks, simultaneous with those in the HXRs. This indicates that the 17 GHz and 35 GHz microwaves are non-thermal {\gyros} emissions. We confirm this assertion from spectral analysis in {\S}~\ref{sec2_3}. Note that the microwave peaks are delayed from the HXR peaks by about 4 s.

\subsection{Images}\label{sec2_2}
\begin{figure}[htbp]
\centering
\plotone{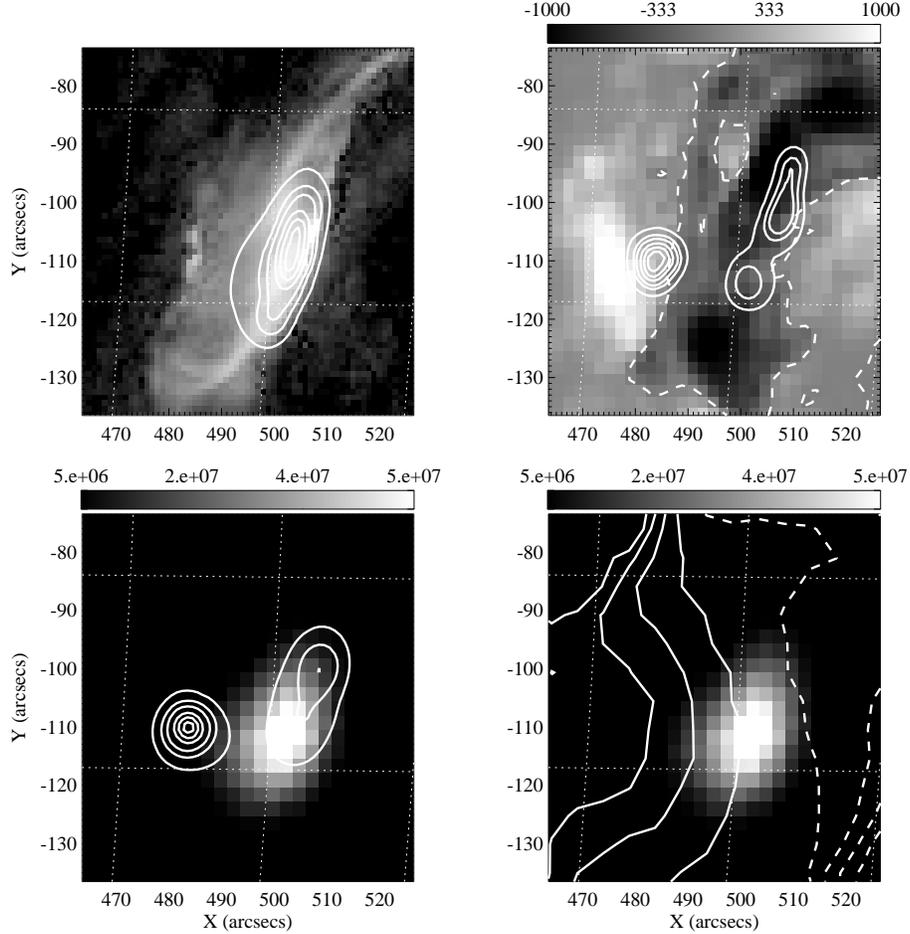}
 \caption{Spatial distribution of the emissions during the impulsive phase of the 2003 May 29 flare (around 01:04:30 UT). The {\it top left} map is the TRACE 195 {\AA} image overlaid by the RHESSI 10-20 keV contours. The {\it top right} map is the SOHO/MDI magnetogram (in units of Gauss) overlaid by the RHESSI 50-100 keV contours. The dashed lines denote magnetic neutral lines. The {\it bottom left} map is the NoRH 34 GHz brightness temperature (in units of kelvin) image overlaid by the RHESSI 100-200 keV contours. The RHESSI images are reconstructed using the PIXON algorithm. Contour levels are 10\%, 20\%, 30\%, 50\%, 70\%, and 90\% of the peak intensity in each image. Thick and dashed contours overlaid on the NoRH 34 GHz brightness temperature image in the {\it bottom right} map are degree of right- and left-circular polarization at 17 GHz, respectively. Contour levels are 10\%, 30\%, 50\%, and 70\%. The angular resolution (beam size) of the NoRH 34 GHz map is $\sim 9''$. The dotted lines denote heliographic grids in $2^{\circ}$ increments.}
\label{maps}
\end{figure}

% Fig. \ref{maps} shows the spatial distribution of the emissions around 01:04 UT. The left map is the TRACE 195 {\AA} image overlaid by the RHESSI 10-20 keV contours. The middle map is the SOHO/MDI magnetogram overlaid by the RHESSI 50-60 keV contours. The right map is the NoRH 34 GHz brightness temperature image overlaid by the RHESSI 100-300 keV contours. The RHESSI images are reconstructed with the CLEAN algorithm \citep{2002SoPh..210...61H} using collimators 3-7, and their accumulation time is 120 seconds (01:03 - 01:05 UT).%  The RHESSI images show the asymmetric HXR sources that the east (west) one is strong (weak).
Figure \ref{maps} shows the spatial distribution of the emissions around 01:04:30 UT. The TRACE 195 {\AA} image ({\it top left}), the SOHO/MDI magnetogram ({\it top right}), and the NoRH 34 GHz brightness temperature image ({\it bottom left}) are overlaid by the RHESSI 10-20 keV, 50-100 keV, and 100-200 keV contours, respectively. The RHESSI images are reconstructed with the PIXON algorithm \citep{2002SoPh..210...61H} using collimators 3-9 with an accumulation time of 60 s (01:04:00 - 01:05:00 UT). The map of the degree of polarization at 17 GHz ({\it bottom right}) is overlaid on the NoRH 34 GHz brightness temperature image.

The TRACE 195 {\AA} image (Fig. \ref{maps}; {\it top left}) shows a typical two-ribbon and arcade structure. The HXR (10-20 keV) source is co-spatial with the brightest region in 195 {\AA}, indicating that the 10-20 keV HXR emission is thermal {\bremss} from the coronal plasma in the flare loop.% Because 195 {\AA} extreme ultraviolet are mainly emitted from the coronal plasma, this HXR emission will correspond to the thermal {\bremss} from the coronal plasma in the flare loop.

The RHESSI maps at 50-100 keV and 100-200 keV (Fig. \ref{maps}; {\it top right} and {\it bottom left}) show double sources located at regions of opposite magnetic polarity (Fig. \ref{maps}; {\it top right}) with one of the sources lying at the edge of the bright region in 195 {\AA} (Fig. \ref{maps}; {\it top left}). Therefore, the HXRs in these energy ranges must be emitted near the footpoints of the loop. Note that the eastern HXR source is brighter than the western one. The magnetic field strength at the eastern source ($\sim +410$ Gauss) is weaker than that of the western one ($\sim -510$ Gauss). Since a stronger HXR source indicates a more efficient electron precipitation, the spatial relationship between the HXR sources and the magnetic field strength can be interpreted as the result of magnetic mirroring of the HXR emitting electrons \citep{1994PhDT.......335S}. % the asymmetric HXR sources that the east (west) one is strong (weak). These are non-thermal footpoint HXR sources because they are located at the edges of the TRACE 195 {\AA} bright region with opposite magnetic polarities. Average intensities of the longitudinal magnetic field at the east and west HXR sources are $\sim +410$ G and $\sim -510$ G, respectively. Thus the east (west) HXR source is located at the region with a weaker (stronger) magnetic field. This is consistent with the magnetic mirroring effect on the HXR emitting electrons \citep{1994PhDT.......335S}.

The microwave source is located between the footpoint HXR sources (Fig. \ref{maps}; {\it bottom left}). It is close to the coronal HXR (10-20 keV) source. The microwave source is also located at the region with weaker degree of polarization (Fig. \ref{maps}; {\it bottom right}). We confirm from the spectral analysis ({\S}~\ref{sec2_3}) that the microwave emissions above 17 GHz are optically-thin non-thermal {\gyros} emissions. Therefore, both the polarization information and the configuration of the longitudinal magnetic field indicate that the magnetic field at the microwave source is quasi-perpendicular to the line of sight, which thus corresponds to the loop-top. Such a spatial distribution of the HXR and microwave emissions can be explained by the TPP model, if the microwaves are emitted by electrons trapped in the loop-top and the HXRs are emitted by electrons precipitating into the footpoints. % We will study this model later.

\subsection{Spectra}\label{sec2_3}
% One of the most direct, and important information on flare non-thermal electrons extracted from the observations is the spectrum. The index, the spatial distribution, and  the temporal evolution of the spectrum must provide us a best chance to discuss physics of flare non-thermal electrons. In the previous studies, the temporal evolution of the (HXR) spectrum has been less considered in detail, probably due to instrumental limitations. However, the temporally-resolved analysis of non-thermal emissions is very important because non-thermal emissions and thus non-thermal electrons are the most ``time-varying'' objects in solar flares. RHESSI enables us, for the first time, to deal with an accurate, temporally-resolved flare HXR spectrum with energy above $\sim 100$ keV.

\begin{figure}[htbp]
\centering
\plotone{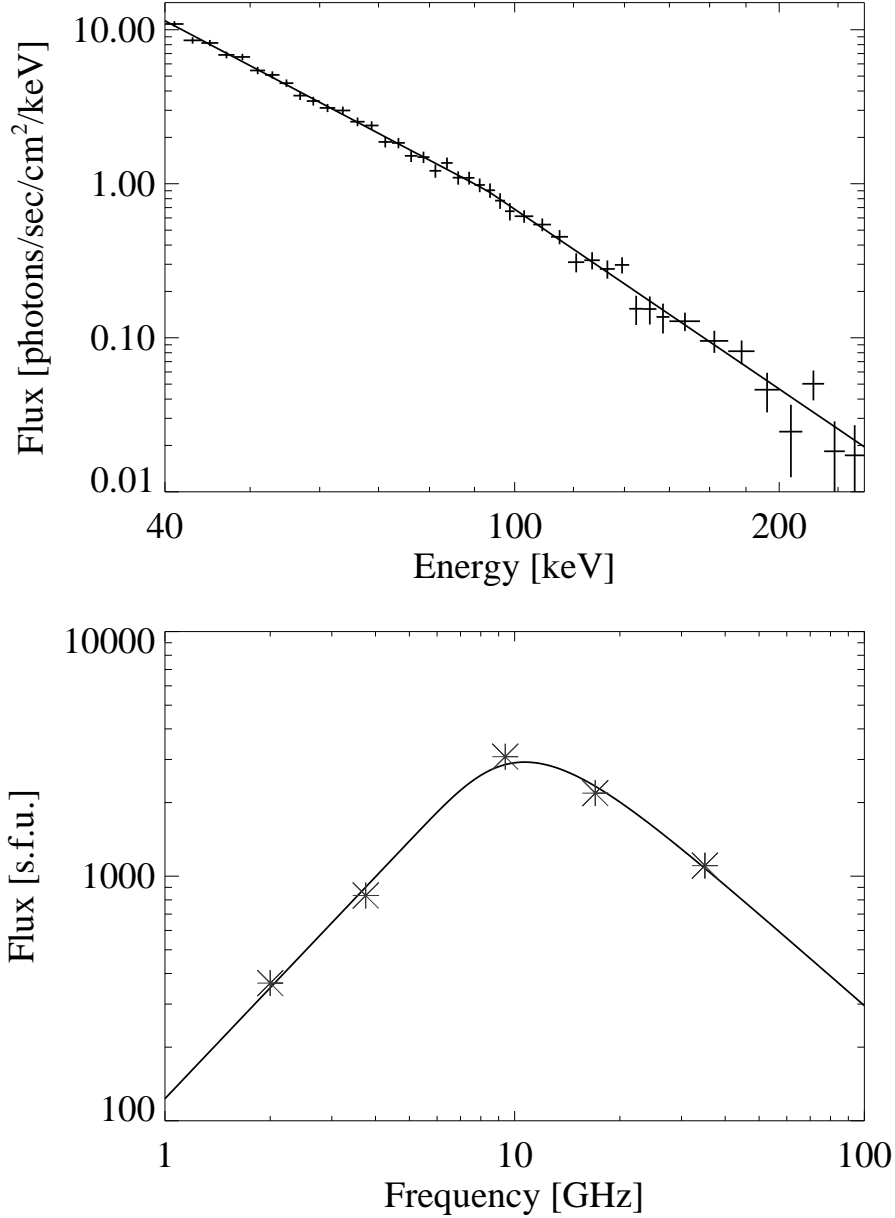}
\caption{HXR and microwave spectra taken with RHESSI and NoRP during the 2003 May 29 flare. The {\it upper} plot shows the HXR photon spectrum (points with error bars) and the fitted double power-law function (solid line) during 01:04:28 - 01:04:32 UT. Values of $\gamma_{\rm L}$, $\gamma_{\rm H}$, and $\epsilon_{\rm b}$ (keV) are determined to be $\{3.01 \pm 0.001, 3.88 \pm 0.003, 93.0 \pm 0.1\}$. The {\it lower} plot shows the microwave spectrum at five frequencies of 2, 3.75, 9.4, 17, and 35 GHz (asterisks) and the fitted model described by equation (\ref{func_gyros}) (solid line) at 01:04:29 UT. Values of $\alpha$ and $\nu_{\rm turnover}$ (GHz) are determined to be $\{1.25, 11.1\}$} 
\label{ex_spectrum}
\end{figure}

We analyzed the temporally-resolved (but spatially-unresolved) spectra during the impulsive phase of the flare. We fit the RHESSI 40-250 keV spectrum at each time interval with a double power-law function of the form
% \begin{eqnarray}
% f(\epsilon,t) && = \nonumber \\
% && \left\{ 
% \begin{array}{ll}
% a(t) \epsilon^{- \gammat{L}{}}, & {\rm if} \; \epsilon \leq \epsilon_{\rm b}(t), \\
% \left[a(t) \epsilon_{\rm b}(t)^{\gammat{H}{} - \gammat{L}{}}\right] \epsilon^{- \gammat{H}{}}, & {\rm if} \; \epsilon > \epsilon_{\rm b}(t), \label{dpowl} \\
% \end{array}
% \right.
% \end{eqnarray}
\begin{eqnarray}
f(\epsilon,t) =
 \left\{ 
\begin{array}{ll}
a(t) \epsilon^{- \gammat{L}{}}, & \; {\rm if} \; \epsilon \leq \epsilon_{\rm b}(t), \\
b(t) \epsilon^{- \gammat{H}{}}, & \; {\rm if} \; \epsilon > \epsilon_{\rm b}(t), \label{dpowl} \\
\end{array}
\right.
\end{eqnarray}
where $\epsilon$ is the photon energy, $\gammat{L}{}$ and $\gammat{H}{}$ are the spectral indices of the lower- and higher-energy parts, $\epsilon_{\rm b}(t)$ is the break energy, and $b(t) = a(t) \epsilon_{\rm b}(t)^{\gammat{H}{} - \gammat{L}{}}$, respectively. The upper panel of Figure \ref{ex_spectrum} shows an example RHESSI energy spectrum and its fitting result. We chose energy bins of 2 keV from 40 to 60 keV, 2.5 keV from 60 to 100 keV, 5 keV from 100 to 150 keV, and 12.5 keV beyond 150 keV, and a temporal resolution of 4 s (approximately equal to the RHESSI rotation period). We used the front segments of detectors \#3, 4, and 8, which have the best energy resolution below $\sim 100$ keV \citep{2002SoPh..210...33S}. For convenience of analysis, the range of $\epsilon_{\rm b}(t)$ was limited to 70 to 130 keV. 

% On the other hand, we can obtain the temporally-resolved microwave spectral index from the NoRP data. We confirm that the microwave turnover frequency is $\sim$10 GHz during the impulsive phase. Then the NoRP data at 17 GHz and 35 GHz are utilized to derive the spectral index of the microwave in the optically thin part \citep{2000ApJ...545.1116S}.
 We also obtained the temporally-resolved microwave spectral index from the NoRP data. After integrating it by 2 s to improve the statistics, we fit the NoRP spectrum taken with five frequencies of 2, 3.75, 9.4, 17, and 35 GHz at each time interval with a generic function \citep{2000ApJ...545.1116S},
\begin{eqnarray}
g(\nu,t) &=& a_{1}(t) \nu^{a_{2}(t)} \left[ 1 - \exp(-a_{3}(t) \nu^{-a_{4}(t)})\right] \\
&\simeq& \left\{
\begin{array}{ll}
a_{1}(t) \nu^{a_{2}(t)}, & \; {\rm if} \; \nu \ll \nu_{\rm turnover}(t), \\
a_{1}(t)a_{3}(t) \nu^{-[a_{4}(t) - a_{2}(t)]} , & \; {\rm if} \; \nu \gg \nu_{\rm turnover}(t), \nonumber \\
\end{array}
\right.
 \label{func_gyros}
\end{eqnarray}
where $\nu$  is the frequency. We obtained the best-fit spectral index of the microwave flux density in the optically thin (higher frequency) regime, $\alpha(t) = a_{4}(t) - a_{2}(t)$ (positive value), as well as the turnover frequency $\nu_{\rm turnover}(t)$. The lower panel of Figure \ref{ex_spectrum} shows an example NoRP microwave spectrum and its fitting result. We confirm that $\nu_{\rm turnover}(t)$ is less than 17 GHz during the impulsive phase, indicating that the microwave emissions above 17 GHz are certainly optically thin, non-thermal {\gyros} emission.

\begin{figure}[htbp]
\centering
\plotone{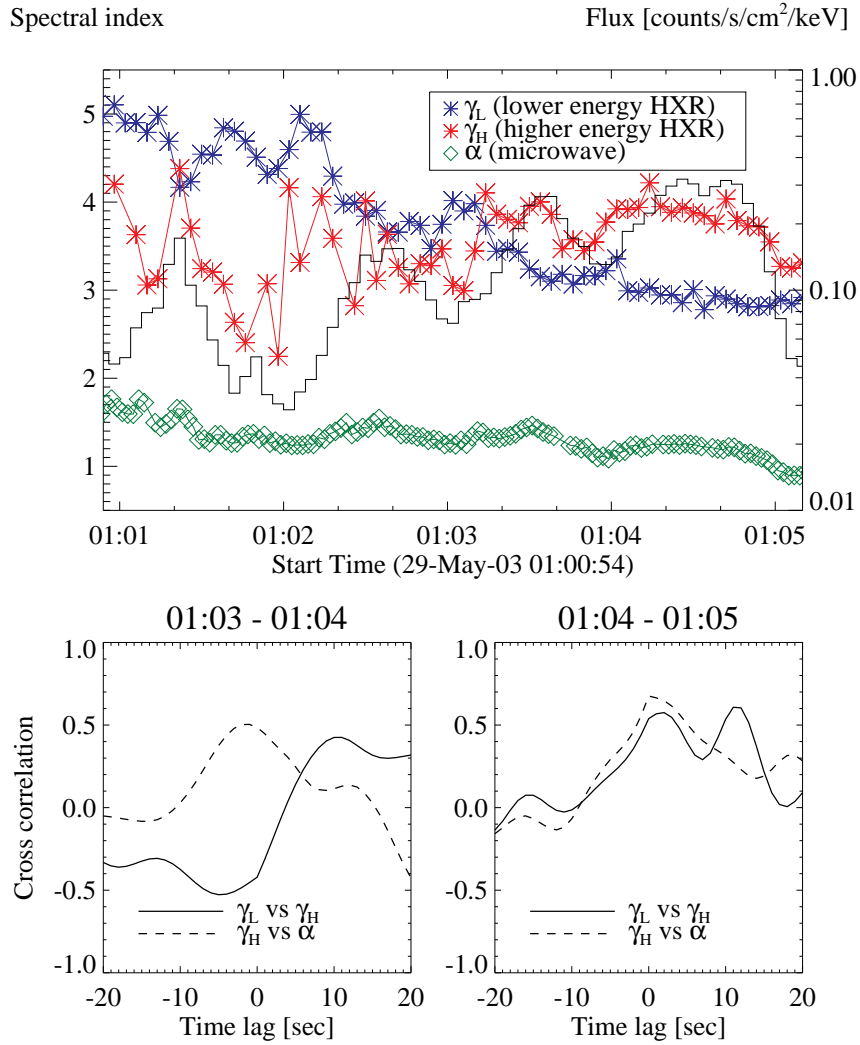}
\caption{{\it Upper}: Time profiles of the spectral indices of the non-thermal emissions during the impulsive phase of the 2003 May 29 flare. The blue, red, and green symbols are $\gammat{L}{obs}$, $\gammat{H}{obs}$, and $\alphat{obs}$, respectively. The break energy is $\sim 100$ keV. The solid line denotes the RHESSI count flux at 50-100 keV. Note that some data which have large uncertainty are omitted in the plot. {\it Lower}: Cross-correlation functions of two time profiles of the spectral indices as a function of time lag, during 01:03 - 01:04 UT ({\it left}) and 01:04 - 01:05 UT ({\it right}). The solid lines are cross-correlation of $\gammat{L}{obs}$ and $\gammat{H}{obs}$, and the dashed lines are that of $\gammat{H}{obs}$ and $\alphat{obs}$, respectively. }
\label{spec}
\end{figure}

Figure \ref{spec} ({\it upper}) shows the time profiles of the spectral indices of the non-thermal emissions. The blue and red asterisks are the spectral indices of the lower-energy ($\lsim 100$ keV) and higher-energy ($\gsim 100$ keV) HXRs (hereafter $\gammat{L}{obs}$ and $\gammat{H}{obs}$), and the green diamonds are the spectral indices of the microwaves in the optically thin regime (hereafter $\alphat{obs}$), respectively.
%  The fitted spectral indices which coefficients of variation exceed 0.1 are plotted by small symbol. The spectral indices of the higher-energy HXR before 01:02:20 UT have large uncertainties because the HXR ($\gsim 100$ keV) intensity is not sufficient. We do not use the fitting results in this period for following discussions.

% Behavior of the spectra found in Fig. \ref{spec} are as follows.
 The time profile of $\gammat{L}{obs}$ shows the so-called {\it soft-hard-soft} behavior for each spike except the last. For example, $\gammat{L}{obs}$ is $\sim 4.8$ at 01:02:10 UT, becomes hard ($\sim 3.7$) at 01:02:40 UT, and softens again ($\sim 4.0$) at 01:03:10 UT. In addition, $\gammat{L}{obs}$ also shows the so-called {\it soft-hard-harder} behavior during the entire course of the impulsive phase. However, the time profile of $\gammat{H}{obs}$ behaves differently from that of $\gammat{L}{obs}$. Neither the {\it soft-hard-soft} behavior nor the {\it soft-hard-harder} behavior can be seen in the time profile of $\gammat{H}{obs}$. The values of the microwave spectral index are quite smaller than those of the HXR spectral indices. This implies that the inferred energy distribution of the microwave emitting electrons is harder than that of the HXR emitting ones. We will carry out a spectral analysis in {\S} \ref{sec4} to confirm whether the energy distribution of the microwave emitting electrons is actually harder than that of the HXR emitting ones. % The microwave spectral index is quite harder than the HXR spectral indices, which indicates that the inferred energy distribution of the microwave emitting electrons is harder than that of the HXR emitting ones.
 %  If we follow \cite{2000ApJ...545.1116S}, this means that the spectral index of the microwave-emitting electrons is harder by $\sim 2$ than that of the HXR-emitting electrons.

In addition to this, we find that the time profile of $\gammat{H}{obs}$ is similar to that of $\alphat{obs}$, although the absolute values of their spectral indices differ by $\sim 2$. We also find that the time profile of $\gammat{H}{obs}$ (and $\alphat{obs}$) is delayed from that of $\gammat{L}{obs}$. This tendency is especially seen during 01:03 - 01:05 UT. The cross-correlation functions of the spline-interpolated time profiles of the spectral indices are shown in the lower panels of Figure \ref{spec}. We find: (1) the time profile of $\alphat{obs}$ and that of $\gammat{H}{obs}$ show a peak correlation without a time delay (within a temporal resolution of 4 s); and (2) the time profile of $\gammat{H}{obs}$ is delayed by about 10 s from that of $\gammat{L}{obs}$. The similarity of the time profiles between $\gammat{H}{obs}$ and $\alphat{obs}$ indicates that the higher-energy HXR emitting electrons and the microwave emitting electrons are from the same population and in a similar energy range. The delay of the time profile of $\gammat{H}{obs}$ from that of $\gammat{L}{obs}$ may be interpreted as an electron energy-dependent time delay \citep[e.g.,][]{1997ApJ...487..936A}.% , like as the energy-dependent time delay of the HXR lightcurves reported by e.g., \cite{1997ApJ...487..936A}.

\section{A model for electron transport}\label{sec3}
 We modeled the observed spectral behavior in the 2003 May 29 flare in terms of TPP. The TPP model has been implemented in various ways, analytic \citep{1976MNRAS.176...15M,1983A&A...119..297M,1986A&A...163..239M,1988A&A...194..279M,1991A&A...242..256M,1981ApJ...251..781L,1988ApJ...327..405L} and numerical approaches \citep{1990A&A...234..487M,1990A&A...230..213M,1990ApJ...354..726H,1991A&A...251..693M,1998ApJ...505..418F,2000ApJ...543..457L} to treat electron transport in the flare loop. We follow the approach made by \cite{1990A&A...230..213M} that allows an explicit treatment of electron pitch-angle diffusion in a spatially-homogeneous magnetic loop.

% We attempt to interpret the observed spectral behavior in the 2003 May 29 flare, in terms of a single population electron transport, the TPP.
% The TPP model describes time evolution of the spectral, spatial, and pitch-angle distribution of the electrons trapped in a converging magnetic loop, under the effects of (1) (unspecified) electron injection into the trap region, (2) electrons energy and pitch-angle changes due to probable physical processes such as the Coulomb collision and wave-particle interactions, and (3) electrons in the loss cone escape from the trap region. Its analytic treatment and application to the solar flare emissions were first presented by \cite{1976MNRAS.176...15M} and many authors have followed it \citep{1998ApJ...505..418F,1983A&A...119..297M,1986A&A...163..239M,1988A&A...194..279M,1991A&A...242..256M,1991A&A...251..693M,1990A&A...234..487M,1990A&A...230..213M,2005ApJ...619.1153M,1981ApJ...251..781L,2000ApJ...543..457L,1988ApJ...327..405L,1990ApJ...354..735L}. In \cite{1976MNRAS.176...15M} and \cite{1986A&A...163..239M}, they have derived analytic solutions of the electrons continuity equation in energy. A general description of the gyro-averaged {\FP} equation has been presented in e.g., \cite{1988ApJ...327..405L,1990ApJ...354..735L}. A numerical treatment of the gyro-averaged {\FP} equation and its application to various physical circumstances are summarized by \cite{1990ApJ...354..726H}.

\subsection{Basic equations}\label{sec3_1}
We treat the spatially-homogeneous, gyro-averaged {\FP} equation following \cite{1990A&A...230..213M}:
\begin{eqnarray}
\pdif{N}{t} + \pdif{}{E}\left(\dot{E} N\right) + F(E,\mu,t) = \pdif{}{\mu}\left(D_{\mu \mu} \pdif{N}{\mu} \right) + Q(E,\mu,t). \label{fp_eq}
% \pdif{N}{t} + \pdif{}{E}\left(\dot{E} N\right) - \pdif{}{\mu}\left(D_{\mu \mu} \pdif{N}{\mu} \right) = Q(E,\mu,t) - F(E,\mu,t). \label{fp_eq}
\end{eqnarray}
Here, $N(E,\mu,t)$ is the trapped electron distribution (number of electrons per unit energy per unit pitch-angle cosine), $Q(E,\mu,t)$ is the electron flux (number of electrons per unit energy per unit pitch-angle cosine per unit time) injected into the trap region, $F(E,\mu,t)$ is the electron flux precipitating into the footpoints, $\dot{E}$ and $D_{\mu \mu}$ are the Coulomb energy loss rate and pitch-angle diffusion coefficient, $E = \Gamma - 1$ is the kinetic energy in units of the electron rest mass energy $m_{\rm e} {c}^2$, $\Gamma$ is the Lorentz factor, $m_{\rm e}$ is the electron mass, $c$ is the speed of light, and $\mu$ is the pitch-angle cosine, respectively.

We adopt the Coulomb energy loss rate and pitch-angle diffusion coefficient for a fully ionized plasma given by \cite{1981ApJ...251..781L},
\begin{eqnarray}
&& \dot{E} = - K n / \beta \equiv - \nu_{\rm E} E, \;\;\; (K = 4 \pi c r_{0}^{2} \ln \Lambda), \label{edot} \\
&& D_{\mu \mu} = \fr{K n}{\beta^{3} \Gamma^{2}} (1-\mu^{2}) = \frac{\nu_{\rm E}}{(E+2)}(1-\mu^{2}) ,\label{dmumu}
\end{eqnarray}
where $r_{0} = 2.82 \times 10^{-13} \; {\rm cm}$ is the classical electron radius, $n$ is the ambient plasma number density, $\ln \Lambda \simeq 25$ is the Coulomb logarithm for the typical solar coronal condition, $\nu_{\rm E} \equiv K n/(\beta E)$ is the Coulomb collision frequency, and $\beta = \mart{1 - \Gamma^{-2}}$. The ambient plasma number density and the Coulomb logarithm are treated as constant in this paper. We neglect other Coulomb diffusion coefficients such as $D_{E E}$ and $D_{\mu E}$ which are smaller than $D_{\mu \mu}$ by a factor of order $\ln \Lambda$ \citep{1990ApJ...354..726H}. Here, any other physics of electron kinematics such as wave-particle interactions are ignored for simplicity.

For the precipitating electron flux $F(E,\mu,t)$, we again follow \cite{1990A&A...234..487M,1990A&A...230..213M},
\begin{eqnarray}
F(E,\mu,t) &=& \fr{H(\abs{\mu} - \mu_{\rm c})}{L/ \left[\abs{\mu} v(E)\right]} N(E,\mu,t) \nonumber \\
&\equiv& \fr{H(\abs{\mu} - \mu_{\rm c})}{\tau_{\rm e}(E,\mu)} N(E,\mu,t) \label{precipitate},\end{eqnarray}
where $L$ is the characteristic scale length (i.e., loop length), $H$ is the Heaviside step function, $\mu_{\rm c}$ is the loss cone angle cosine, $v(E) = c\beta = c \left[1 - \left(1 + E\right)^{-2}\right]^{1/2}$ is the velocity of an individual electron, and $\tau_{\rm e}(E,\mu)$ is the electron loop-transit time along a magnetic field line, respectively. Here we assume a symmetrical magnetic loop with abrupt increases of magnetic field intensity and ambient plasma number density below the mirror points near the footpoints, and we thus assume symmetrical precipitation in each half of $\mu$ space ($\mu < 0$ and $\mu > 0$). This form of equation (\ref{precipitate}) is for mathematical convenience \citep{1988A&A...194..279M}.

We take the parameters as follows: $L = 3 \times 10^{9} \; {\rm cm}$, $n = 3 \times 10^{10} \; {\rm cm^{-3}}$, and $\mu_{\rm c} = 0.7$. We adopt the value for the characteristic scale length $L$ based on the distance of the HXR sources in the 2003 May 29 flare, assuming a semi-circular shape for the loop. The number density $n$ in our model is slightly smaller than the observed value of $(6-8) \times 10^{10} \; {\rm cm^{-3}}$ derived from the GOES observation during 01:03 - 01:05 UT (assuming a volume of $L^3$ and a filling factor of unity). We assume that electrons are trapped in an outer loop with a lower density than the brightest SXR loop. In the outer loop, energy dissipation (electron bombardment) at the footpoints and the resultant filling with evaporated chromospheric plasma have not yet occurred, whereas they have already occurred in the inner SXR loop. % Our choice of $n$ is based on their idea. However, the topology of electron trap that the trap region is identical to the SXR loop is also presumable. We will discuss the calculation results for several values of $n$.
The value for the loss cone angle cosine $\mu_{\rm c}$ corresponds to a magnetic mirror ratio of 2, that is, the ratio of magnetic field strength at the footpoint to that at the trap region. We adopt this value based on the statistical analysis of flare data taken with {\it CGRO} and {\Yohkoh} by \cite{1998ApJ...502..468A,1999ApJ...517..977A}. They estimated the fraction of directly-precipitating electrons to trap-precipitating ones and derived a magnetic mirror ratio of $1.2 - 3$.

\subsection{Time-dependent injection flux}\label{sec3_2}
We give the time-dependent, single power-law electron injection flux $Q(E,\mu,t)$,
\begin{eqnarray}
Q(E,\mu,t) = A(t) \left(\frac{E}{E_{0}} \right)^{-\deltat{in}} \phi(\mu), \label{injection}
\end{eqnarray}
where we adopt a {\it pivot point energy} \citep{2006A&A...458..641G} $E_{0}$ of 0.098 (= 50 keV), and
\begin{eqnarray}
A(t) &=& \exp\left[-\left(\fr{t-30}{25}\right)^{2}\right] + \exp\left[-\left(\fr{t-90}{25}\right)^{2}\right] \label{inj_a} \\
% A(t) &=& \left\{\exp\left[{- \left(\fr{t-30}{25}\right)^{2} }\right] \right.  \nonumber \\
% && + \left.\exp\left[{- \left(\fr{t-90}{25}\right)^{2} }\right]\right\}, \label{inj_a} \\
\deltat{in} &=& 4.5 + \cos^{2} \left(\fr{\pi t}{60} \right), \label{inj_d}
\end{eqnarray}
for $0 \leq t \leq 120$ sec. Equations (\ref{inj_a}) and (\ref{inj_d}) describe double peaks electron injection with a {\it soft-hard-soft} spectrum. We adopt this form for the electron injection because the HXR spectrum in the lower-energy regime observed in the 2003 May 29 flare shows the {\it soft-hard-soft} behavior.
% Such electron injection is expected to produce the double-peaks HXR with {\it soft-hard-soft} spectrum, like as that shown in Fig. \ref{spec} (solid line and blue asterisks during 01:03 - 01:05 UT).

The remaining term of $\phi(\mu)$ gives the (time- and energy-independent) pitch-angle distribution of the injection flux. We perform calculations for two cases of the pitch-angle distribution: the pancake distribution, $\phi(\mu) = \exp[- (\mu / 0.5)^{2}]$, and the isotropic distribution, $\phi(\mu) = {\rm const}$.

% Using eqs. (\ref{edot}) - (\ref{inj_d}), we numerically solve eq. (\ref{fp_eq}) and obtain $N(E,\mu,t)$ and $F(E,\mu,t)$. For a spectral analysis, we use the $\mu$-integrated $N(E,\mu,t)$ and $F(E,\mu,t)$ (hereafter $N_{\mu}(E,t)$ and $F_{\mu}(E,t)$, respectively).%  In the next section, we will show the time evolution of their spectra.

Using equations (\ref{edot}) - (\ref{inj_d}), we numerically solve equation (\ref{fp_eq}). We apply a finite difference method with operator splitting. The differential operators in equation (\ref{fp_eq}) are split into two terms: the diffusion term and the remaining terms. We solve the diffusion term by using the Crank-Nicholson method with central difference, accurate to second order in time. We employ the symmetric boundary condition in $\mu$ space which satisfies total number conservation. We set 97 grid points in $\mu$ space by the following manner,
\begin{eqnarray}
&& \mu_{0} = 0, \;\; \Delta\mu = 0.016, \nonumber\\
&& \mu_{n} = \mu_{n-1} + \Delta\mu(1-\mu_{n-1}^{2}), \;\; (n=1,2,\cdots,95), \label{mu_grids} \\
&& \mu_{96} = 1, \nonumber
\end{eqnarray}
giving coarse grids for smaller $\mu$ and fine grids around the loss cone.
For the remaining term, we use an analytic solution given by the method of characteristics \citep[see, e.g.,][]{1985Ap&SS.116..377C,1986A&A...163..239M}. For the necessary interpolation at the intermediate location between the grid points in energy space, we use a single power-law function. %  In energy space, we use a single power-law function to interpolate $N(E,\mu,t)$ between neighboring grids, and to extrapolate it from the two nearest grids at the boundary. 
We set 256 grid points in energy space, logarithmically-spaced from 50 to 5000 keV.

% Consequently we describe the time evolution of $N(E,\mu,t)$ and $F(E,\mu,t)$. For a spectral analysis, we introduce the $\mu$-integrated variables,
% \begin{eqnarray}
% \left[ N_{\mu}(E,t), F_{\mu}(E,t) \right]
% =  \inty{-1}{1} d \mu \left[ N(E,\mu,t), F(E,\mu,t) \right]. \label{mu_int_val}
% \end{eqnarray}

\section{Calculation results and discussion}\label{sec4}
\begin{figure}[htbp]
\centering
%\epsscale{0.7}
\plotone{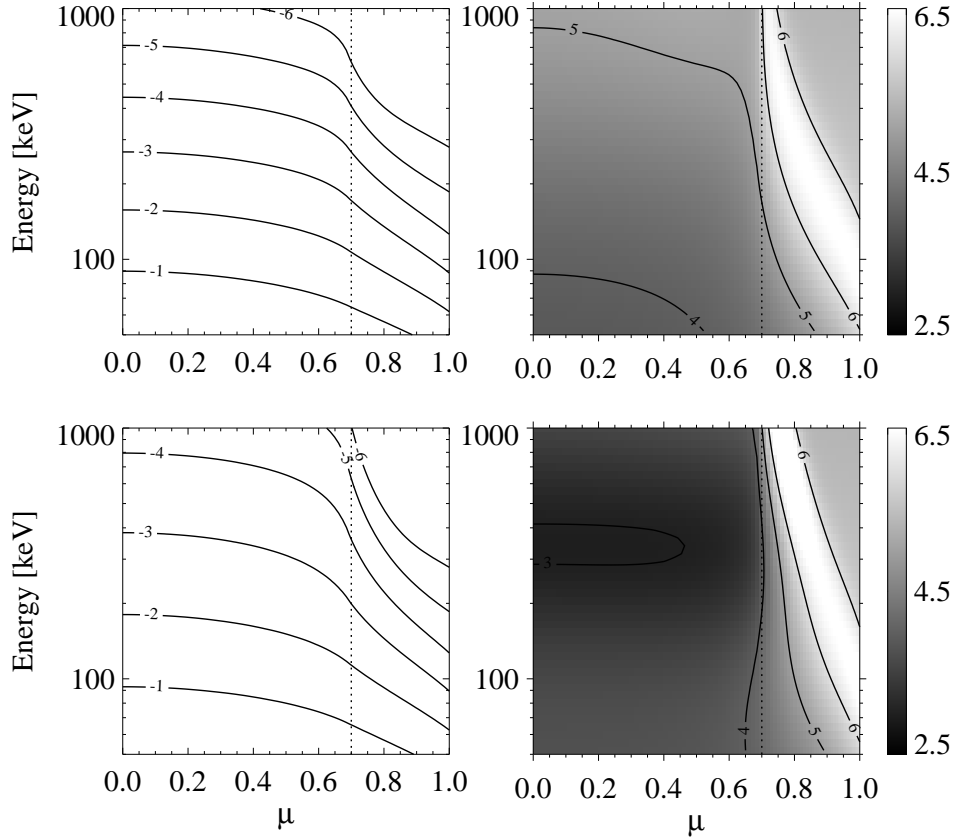}
\caption{Electron distribution in phase space $(E,\mu)$, calculated for a pancake pitch-angle distribution of the injection flux. The {\it top left} panel shows the trapped electron distribution $N(E,\mu,t)$ at a time of $t = 10$ sec. Contour levels are $10^{-6}, 10^{-5}, 10^{-4}, 10^{-3}, 10^{-2}, $ and $10^{-1}$ of the maximum value. The {\it top right} image with contours shows the slope of $N(E,\mu,t)$ in energy, determined from the ratio of $N(E,\mu,t)$ at two adjacent energy grids (eq. (\ref{slope})). Contour levels are 3, 4, 5, and 6. The dotted lines denote the loss cone angle cosine. The {\it bottom} two panels are same as the {\it top} ones, but at a time of $t = 50$ sec.}
\label{data_plt_pan}
\end{figure}

\begin{figure}[htbp]
\centering
%\epsscale{0.7}
\plotone{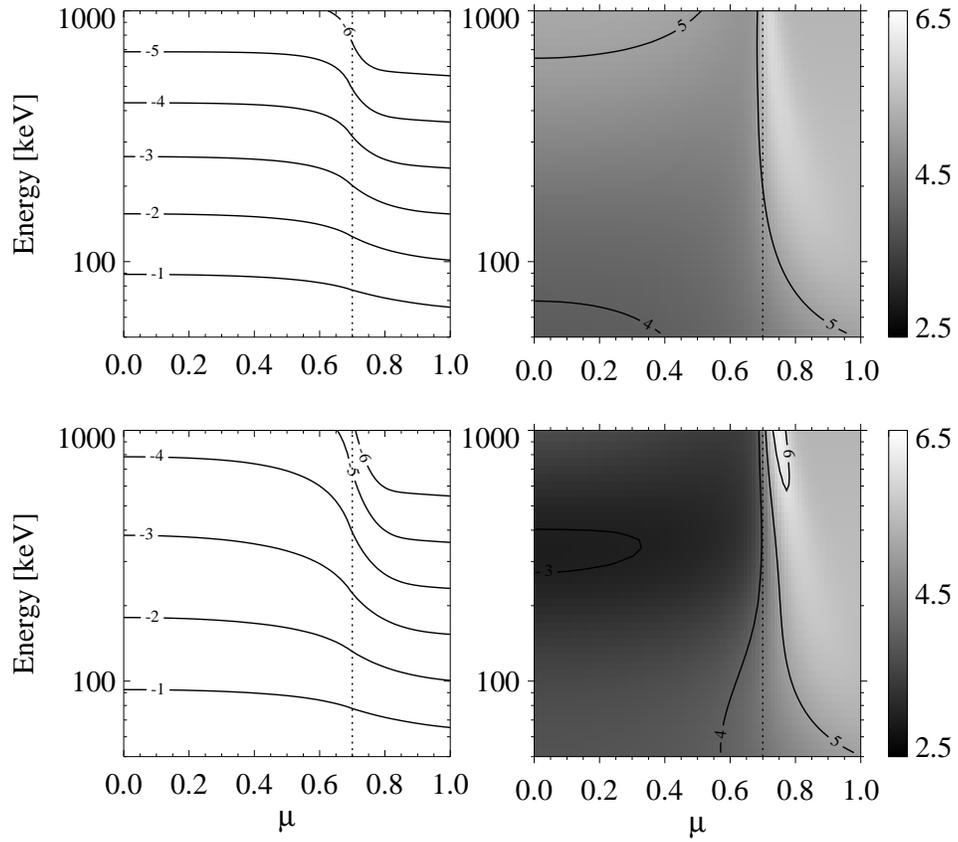}
\caption{Same as Figure \ref{data_plt_pan}, but calculated for an isotropic pitch-angle distribution of the injection flux.}
\label{data_plt_iso}
\end{figure}

In this section, we present our calculation results of the {\FP} equation. Figures \ref{data_plt_pan} and \ref{data_plt_iso} show the electron distribution in phase space $(E,\mu)$, calculated for the pancake and isotropic pitch-angle distributions of the injection, respectively. The left panels show the trapped electron distribution $N(E,\mu,t)$ at selected times of $t=10$ sec ({\it top}) and $t=50$ sec ({\it bottom}). The right panels show the slope $s$ of $N(E,\mu,t)$ in energy at the selected times, derived from the following equation,
\begin{eqnarray}
s_{i} = - \frac{\log \left[ N(E_{i+1},\mu,t)/N(E_{i},\mu,t) \right]}{\log \left( E_{i+1}/E_{i}\right)}, \label{slope}
\end{eqnarray}
where subscript $i$ denotes the grid position in energy space. We can see from these right panels that the electron energy distribution outside the loss cone $(\mu < 0.7)$ is harder in the decay phase ($t=50$ sec) than in the rise phase ($t=10$ sec). The electron energy distribution inside the loss cone $(\mu > 0.7)$, which is related to the precipitating electron flux $F(E,\mu,t)$ by equation (\ref{precipitate}), is softer than that outside the loss cone. The electron distribution inside the loss cone shows different features between the pancake and isotropic cases.

In {\S}~\ref{sec4_1}, we discuss the time evolution of the trapped electron distribution and the precipitating electron flux in energy space by using the double power-law fitting method. Our calculations confirm the electron trap and precipitation regardless of the weak and strong diffusion limits. In {\S}~\ref{sec4_2}, we discuss the time evolution of the non-thermal emissions predicted by the electron model for comparison with the observations.
 
% We first show time evolution of the trapped electron distribution and the precipitating electron flux in {\S} \ref{sec4_1}. For a following spectral analysis of electrons, we introduce the $\mu$-integrated variables,
% \begin{eqnarray}
% N_{\mu}(E,t) &=& \inty{0}{1} d \mu N(E,\mu,t), \nonumber \\
% F_{\mu}(E,t) &=& \inty{0}{1} d \mu F(E,\mu,t) = \inty{\mu_{\rm c}}{1} d \mu \frac{N(E,\mu,t)}{\tau_{\rm e}(E,\mu)}. \label{mu_int_val}
% \end{eqnarray}
% Our calculations confirm the electron trap and precipitation regardless of the weak and strong diffusion limits. In {\S}~\ref{sec4_2} we further show time evolution of the non-thermal emissions predicted by the calculated electrons, and compare them with the observations. Calculation results give some agreements with the observations.
 
\subsection{Evolution of electrons}\label{sec4_1}
\begin{figure}[htbp]
\centering
\epsscale{0.7}
\plotone{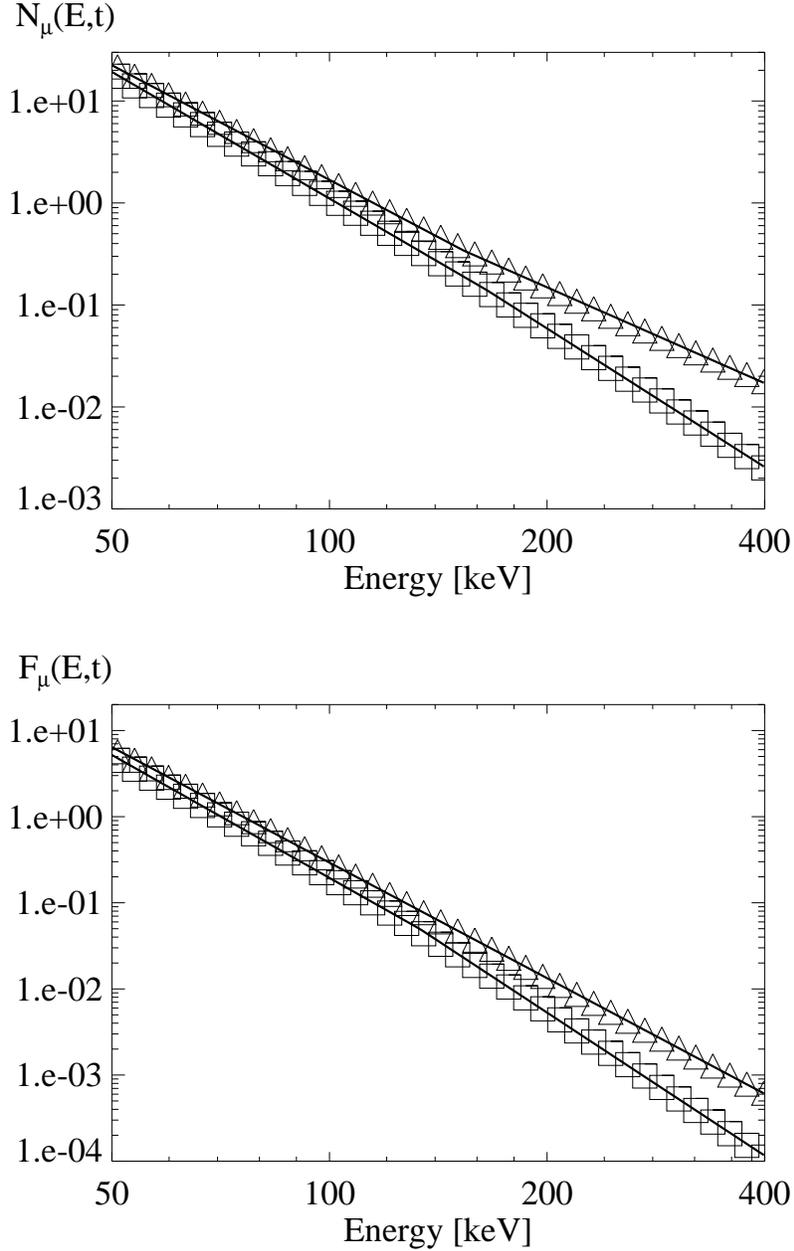}
\caption{Electron energy spectra in the pancake case. The {\it upper} panel shows the energy spectra of the trapped electron distribution $N_{\mu}(E,t)$ at selected times of $t=10$ sec (squares) and $t=50$ sec (triangles) with the fitted double power-law function (solid lines). Values of $\Delta_{\rm L}$, $\Delta_{\rm H}$, and $E_{\rm b}$ (keV) are determined to be $\{4.13 \pm 0.017, 4.51 \pm 0.028, 165 \pm 7.9\}$ at $t=10$ sec, and $\{3.73 \pm 0.019, 3.12 \pm 0.025, 153 \pm 4.7\}$ at $t=50$ sec. The {\it lower} panel shows the energy spectra of the precipitating electron flux $F_{\mu}(E,t)$. Values of $\delta_{\rm L}$, $\delta_{\rm H}$, and $E_{\rm b}$ (keV) are determined to be $\{4.74 \pm 0.023, 5.52 \pm 0.020, 133 \pm 3.1\}$ at $t=10$ sec, and $\{4.45 \pm 0.016, 4.46 \pm 0.018, 100\}$ at $t=50$ sec.}
\label{cal_spec_sample}
\end{figure}

To illustrate the evolution of the trapped and precipitating electrons, we introduce the pitch-angle integrated variables:
\begin{eqnarray}
N_{\mu}(E,t) &=& \inty{0}{1} d \mu N(E,\mu,t), \nonumber \\
F_{\mu}(E,t) &=& \inty{0}{1} d \mu F(E,\mu,t) = \inty{\mu_{\rm c}}{1} d \mu \frac{N(E,\mu,t)}{\tau_{\rm e}(E,\mu)}. \label{mu_int_val}
\end{eqnarray}
Next, we fit the electron energy spectra from 50 to 400 keV at each time by the double power-law function of equation (\ref{dpowl}) (but replacing $[\epsilon,\gamma]$ by $[E,\Delta]$ for $N_{\mu}(E,t)$, and by $[E,\delta]$ for $F_{\mu}(E,t)$). The range of $E_{\rm b}(t)$ is limited to 100 to 200 keV for convenience. An example of our fitting of the electron energy spectra is shown in Figure \ref{cal_spec_sample}. In this way, we obtain the spectral indices $\Deltat{L}$, $\Deltat{H}$, $\deltat{L}$, and $\deltat{H}$.

\begin{figure}[htbp]
\centering
\plotone{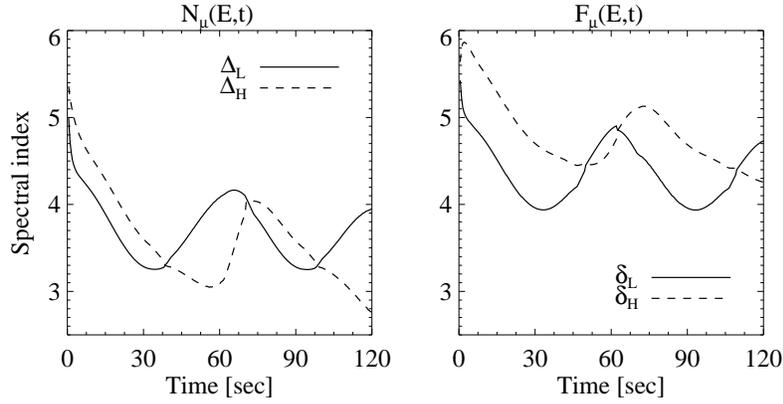}
\caption{Time profiles of the spectral indices of electrons in the pancake case. {\it Left}: Time profiles of $\Deltat{L}$ (lower-energy regime, solid line) and $\Deltat{H}$ (higher-energy regime, dashed line). The break energy is $\sim 170$ keV. {\it Right}: Those of $\deltat{L}$ (lower-energy regime, solid line) and $\deltat{H}$ (higher-energy regime, dashed line). The break energy is $\sim 130$ keV.}
\label{spec_pancake}
\end{figure}

%\subsubsection{Pancake case}\label{sec4_1_1}
Figure \ref{spec_pancake} shows the time profiles of the spectral indices of the electrons in the pancake case. The left panel shows $\Deltat{L}$ (lower-energy regime, solid line) and $\Deltat{H}$ (higher-energy regime, dashed line). The break energy is $\sim 170$ keV. The right panel shows $\deltat{L}$ (lower-energy regime, solid line) and $\deltat{H}$ (higher-energy regime, dashed line). The break energy is $\sim 130$ keV.

% The time profiles of the lower-energy spectral indices ($\Deltat{L}$ and $\deltat{L}$) take their hardest value around $t=30$ and $t=90$ that correspond to the time profile of the spectral index of the injection flux (eq. (\ref{inj_d})). However, the time profiles of the higher-energy spectral indices ($\Deltat{H}$ and $\deltat{H}$) do not show such behavior but delays from that of the lower-energy ones.

 The spectra of $F_{\mu}(E,t)$ are quite softer than those of $N_{\mu}(E,t)$. To understand this in terms of a diffusion regime, we introduce the precipitation rate
\begin{eqnarray}
\nu_{\rm p} = \frac{F_{\mu}}{N_{\mu}} \propto E^{x}. \label{prec_rate}
\end{eqnarray}
In the strong and weak diffusion limits, the precipitation rates are respectively evaluated as (using the non-relativistic expression), 
\begin{eqnarray}
\nu_{\rm p} \propto \left\{
\begin{array}{l}
1/\tau_{\rm e} \propto v \propto E^{0.5}, \;\;\; {\rm (strong)}\\
1/\tau_{\rm d} \sim D_{\mu \mu} \propto E^{-1.5}, \;\;\; {\rm (weak)}
\end{array}
\right. \label{prec_limits}
\end{eqnarray}
that yields $-1.5 \leq x \leq 0.5$. Whether diffusion for an electron is {\it weak} or {\it strong} depends on its energy $E$. This means that $x$ itself is a function of $E$. When $\tau_{\rm d} > \tau_{\rm e}$ (which corresponds to higher $E$), the diffusion is weak and $x$ takes a negative value, and vice versa. In our situation, the energy at which $\tau_{\rm d} \sim \tau_{\rm e}$ is about 20 keV (using $\mu = 0.7$). All of the electrons in our calculation ($\geq 50$ keV) satisfy $\tau_{\rm d} > \tau_{\rm e}$, and so are in the weak diffusion regime. Because $\tau_{\rm d}$ becomes much longer than $\tau_{\rm e}$ for the higher-energy electrons, $x$ approaches its lower limit (weak diffusion limit) of $-1.5$ with increasing $E$. We find that $-1.5 < x < -1.0$ in the high-energy regime and $x \sim -0.7$ in the lower-energy regime from Figure \ref{spec_pancake}. This is consistent with the previous statement, and thus confirms our general treatment of the TPP model in the weak diffusion regime considered by \cite{1976MNRAS.176...15M}.

In the left panel of Figure \ref{spec_pancake}, the time profile of $\Deltat{L}$ shows the {\it soft-hard-soft} behavior in each spike ($0\le t \le60$ and $60\le t \le120$), achieving its hardest values around $t=30$ and $t=90$. However, the time profile of $\Deltat{H}$ does not show such behavior but shows the {\it soft-hard-harder} behavior in each spike. The energy-dependent trap efficiency yields this difference of the temporal variation of the spectral indices between the lower and higher energies. The trapped electrons are lost via Coulomb energy loss and precipitation. The Coulomb energy loss rate $\nu_{\rm E}$ is smaller for the higher-energy electrons. As mentioned above, the weak diffusion yields a precipitation rate that is also smaller for the higher-energy electrons. This means that the escape time scale from the phase space becomes longer for the higher-energy electrons. For the lower-energy electrons, the escape time scale is on the order of 1 sec. This is much shorter than the injection time scale (see eq. (\ref{inj_a})). Therefore, $\Deltat{L}$ reflects the temporal variation of the spectral index of the injection flux $\deltat{in}$ described by equation (\ref{inj_d}). The escape time scale becomes comparable to or longer than the injection time scale for the higher-energy electrons. The higher-energy electrons stay at the trap region and their energy remains high for a longer time. As a result, the spectrum in the higher-energy regime becomes harder in the decay phase of each spike ($30\le t \le60$ and $90\le t \le120$).

\begin{figure}[htbp]
\centering
\plotone{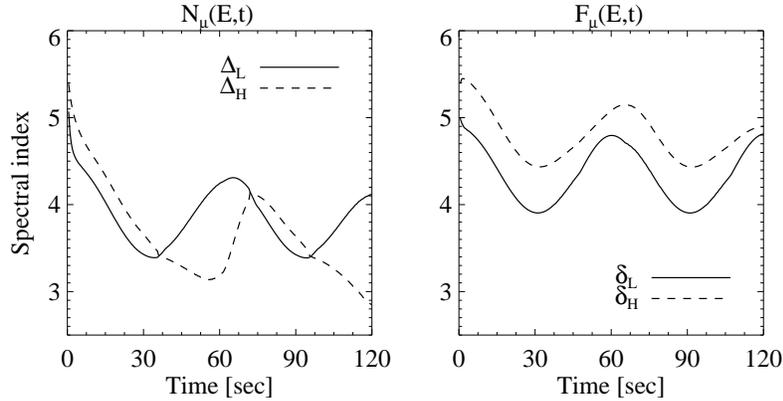}
\caption{Same as Figure \ref{spec_pancake}, but in the isotropic case. Break energies are $\sim 170$ keV ({\it left}) and $\sim 120$ keV ({\it right}).}
\label{spec_iso}
\end{figure}

Figure \ref{spec_iso} shows the time profiles of the spectral indices of the electrons in the isotropic case. The left panel shows $\Deltat{L}$ (lower-energy regime, solid line) and $\Deltat{H}$ (higher-energy regime, dashed line). The break energy is $\sim 170$ keV. The right panel shows $\deltat{L}$ (lower-energy regime, solid line) and $\deltat{H}$ (higher-energy regime, dashed line). The break energy is $\sim 120$ keV.

There is no much difference between the left panels of Figures \ref{spec_pancake} and \ref{spec_iso}, indicating that the evolution of the trapped electron distribution in energy space is almost independent of the pancake or isotropic pitch-angle distribution of the injection. We find a difference between the right panels of Figures \ref{spec_pancake} and \ref{spec_iso}. In both cases, the time profiles of $\deltat{L}$ show the {\it soft-hard-soft} behavior in each spike. In the isotropic case (right panel of Fig. \ref{spec_iso}), the time profile of $\deltat{H}$ shows the {\it soft-hard-soft} behavior same as that of $\deltat{L}$. In the pancake case (right panel of Fig. \ref{spec_pancake}), however, the time profile of $\deltat{H}$ is delayed from that of $\deltat{L}$. We interpret these features in terms of the difference of the injection pitch-angle distribution.

The precipitating electrons consist of two different types: directly-precipitating and trap-precipitating electrons \citep{1998ApJ...502..468A}. Electrons injected with small pitch-angle directly precipitate without being trapped, while those injected with large pitch-angle are trapped once and subsequently precipitate via pitch-angle scattering. When the injected electrons have an isotropic pitch-angle distribution, the precipitating electrons include the directly-precipitating ones as well as the trap-precipitating ones. When the injected electrons have a pancake pitch-angle distribution, on the other hand, almost all of the precipitating electrons are trap-precipitating ones. The right panels of Figure \ref{data_plt_pan}, which show the slope of $N(E,\mu,t)$ in energy in the pancake case, show a ``propagating feature'' (contours around the loss cone) from outside the loss cone to inside the loss cone. This corresponds to the trap-precipitating electrons. The ``propagation speed'' is faster for the lower-energy electrons because it is governed by the pitch-angle diffusion coefficient (eq. (\ref{dmumu})). This feature is less clear in the isotropic case (right panels of Fig. \ref{data_plt_iso}). This is because in the isotropic case the trap-precipitating electrons merge with the directly-precipitating ones.

The directly-precipitating electrons precipitate on a time scale of $\tau_{\rm e}$. The precipitation time scale of the trap-precipitating electrons, $\sim \tau_{\rm d}$, is longer than $\tau_{\rm e}$. A fraction of these two components, which is determined by  the pitch-angle distribution of the injection, determines the precipitation time scale of all of the precipitating electrons. The precipitation time scale is longer in the pancake case than in the isotropic case. For the lower-energy electrons, however, the precipitation time scale is shorter than the injection time scale in both the pancake and isotropic cases. Therefore, $\deltat{L}$ reflects the temporal variation of $\deltat{in}$ regardless of the injection pitch-angle distribution. For the higher-energy electrons, the precipitation time scale becomes longer due to the weak diffusion. It is comparable to or longer than the injection time scale in the pancake case, but not in the isotropic case. Consequently, the time profile of $\deltat{H}$ is delayed from that of $\deltat{in}$ (and thus $\deltat{L}$) only in the pancake case.

\subsection{Evolution of radiation}\label{sec4_2}
In this section, we show the time evolution of the non-thermal emissions predicted by the {\FP} calculation results, and compare them with the observations.
% To do this, we have to estimate the HXR and microwave emissions from the calculated electron distribution.
% First of all, we revisit the images of the 2003 May 29 flare (Fig. \ref{maps} in {\S} \ref{sec2_2}).
 The spatial distribution of the non-thermal emissions in the 2003 May 29 flare shows a loop-top microwave source and double footpoint HXR sources (Fig. \ref{maps}). This supports the interpretation that the trapped electrons $N(E,\mu,t)$ emit microwaves via {\gyros} radiation \citep{1969ApJ...158..753R,1985ARA&A..23..169D} and the precipitating electrons $F(E,\mu,t)$ emit HXRs via thick-target {\bremss} \citep{1971SoPh...18..489B}. We numerically calculate the thick-target HXR intensity at a photon energy $\epsilon$ by%  assuming $F_{\mu}(E,t)$ as the HXR emitter via the isotropic thick-target {\bremss}
\begin{eqnarray}
I(\epsilon,t) &=& \frac{1}{4 \pi R^{2}} \inty{\epsilon}{\infty} dE F_{\mu}(E,t) \inty{\epsilon}{E} dE^{\prime} \fr{n v(E^{\prime}) \sigma_{\rm B}(\epsilon, E^{\prime})}{E^{\prime} \nu_{\rm E}(E^{\prime})}, \label{thick_target}
% I(\epsilon,t) &=& \frac{1}{4 \pi R^{2}} \inty{\epsilon}{\infty} dE F_{\mu}(E,t) \nonumber \\
% && \times \inty{\epsilon}{E} dE^{\prime} \fr{n v(E^{\prime}) \sigma_{\rm B}(\epsilon, E^{\prime})}{E^{\prime} \nu_{\rm E}(E^{\prime})}, \label{thick_target}
\end{eqnarray}
where $\sigma_{\rm B}(\epsilon, E)$ is the direction-integrated {\bremss} cross section given by \cite{1997A&A...326..417H}, $\nu_{\rm E}$ is given in equation (\ref{edot}), and $R = 1\;{\rm AU}$. We use the pitch-angle integrated electron flux because electrons precipitating into the thick-target region are quickly isotropized. 

For the microwave emission, we numerically calculate only the {\gyros} emissivity from the trapped electrons $N(E,\mu,t)$ by using an approximate analytic expression given by \cite{1981ApJ...251..727P} (see Appendix). Although a general description of {\gyros} radiation includes absorption, this approximation is sufficient for our purpose because we discuss the microwave spectral behavior only in the optically-thin regime. We calculate the {\gyros} emissivity in a harmonic range of $10-100$ with a nominal viewing angle $\theta = 75^{\circ}$. When the magnetic field strength at the emission site is on the order of 100 Gauss, the frequency range in our calculation corresponds to the ranges of NoRP and NoRH. 
% Fig. \ref{cal_ltc} shows the HXR lightcurves in the pancake case. The HXR fluxes at 50 keV, 70 keV, 99 keV, 140 keV, and 197 keV in a normalized scale are drawn from top to bottom. The dashed lines denote the peak times of each spike in the HXR flux at 50 keV. In this figure, we can see that the peak times of these HXR lightcurves within 50-200 keV are almost coincident. % The peak time of the HXR lightcurve at 197 keV lags that at 50 keV only by 1.6 seconds.
%  %  In {\S} \ref{sec2_1}, we show that the peak times of the HXR lightcurves within 50-200 keV almost coincide within 2 seconds (Fig. \ref{ltc}).
% Similar tendency is obtained also in the isotropic case.

\subsubsection{Lightcurves}\label{sec4_2_1}
\begin{figure}[htbp]
\centering
\plotone{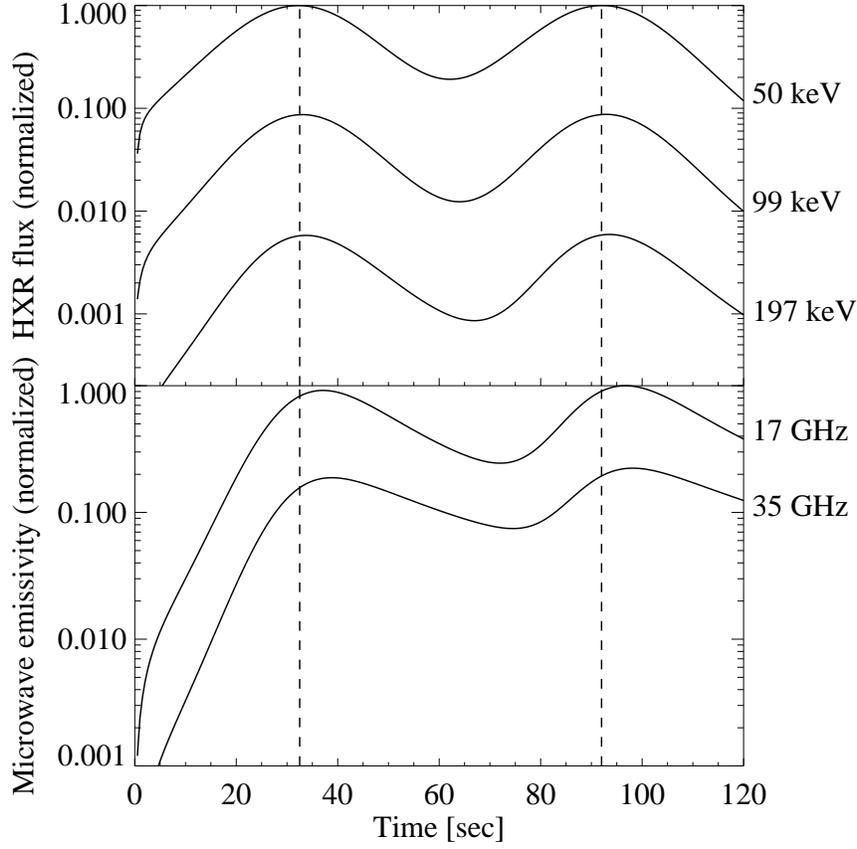}
\caption{Non-thermal emissions predicted by the {\FP} calculation results in the pancake case. The {\it upper} part shows the HXR fluxes at 50 keV, 99 keV, and 197 keV on a normalized scale from top to bottom. The {\it lower} part shows the microwave emissivities at 17 GHz and 35 GHz on a normalize scale from top to bottom. A magnetic field intensity of 300 Gauss and a viewing angle $\theta = 75^{\circ}$ are assumed for the microwave emissivity calculation. The dashed lines denote the peak times of each spike in the HXR 50 keV flux. }
\label{cal_ltc}
\end{figure}

 Figure \ref{cal_ltc} shows the lightcurves of the non-thermal emissions in the pancake case. The upper part shows the HXR fluxes at 50 keV, 99 keV, and 197 keV on a normalized scale. The dashed lines denote the peak times of each spike in the HXR 50 keV flux. We can see that the peaks of the HXR lightcurves within 50-200 keV are almost coincident. The lower part of Figure \ref{cal_ltc} shows the microwave emissivities at 17 GHz and 35 GHz on a normalized scale, assuming a magnetic field strength $B$ at the trap region of 300 Gauss. The peaks of the microwave emissivities are delayed from the HXR flux at 50 keV by about 5 s. These tendencies are also observed in the isotropic case, and are consistent with the non-thermal lightcurves for the 2003 May 29 flare (Fig. \ref{ltc}). 

% For the microwave, we simply assume that $N(E,\mu,t)$ is the emitter via the {\gyros}. However, we have not numerically calculated the microwave emission (the {\gyros} emissivity) yet. To numerically calculate the {\gyros} emissivity properly, we have to execute the calculation up to an enough high energy of electrons ($\gsim 1$ MeV, depends on the harmonic number). This is our future work.

% Next, we present a temporally-resolved spectral analysis on the calculation results. 
\subsubsection{Spectra}\label{sec4_2_2}
To illustrate the spectral variation of the calculated emissions, we fit the calculated HXR spectrum within 50-200 keV at each time with the double power-law function of equation (\ref{dpowl}). The range of $\epsilon_{\rm b}(t)$ is limited to 75 to 125 keV. We also fit the calculated microwave spectrum from 17 to 35 GHz at each time with a single power-law function. 
% Then we again obtain the two spectral indices of the lower- and higher-energy HXR (hereafter $\gammat{L}{FP}$ and $\gammat{H}{FP}$).
% We also deduce the spectral index of the microwave flux density in the optically thin part, $|\alpha(t)|$, from $N_{\mu}(E,t)$%  without calculating the {\gyros} emissivity
% . We adopt the empirical relation between the {\gyros} and the parent electron spectra given by \cite{1985ARA&A..23..169D}:
% \begin{eqnarray}
% |\alphat{FP}| = 1.22 - 0.9 \Deltat{H}, \label{dulk_relation}
% \end{eqnarray}
% where $\Deltat{H}$ is the spectral index of the higher-energy part of $N_{\mu}(E,t)$, defined in \S~\ref{sec4_1}.
% where $\alpha(t)$ is the spectral index of the microwave in the optically thin part, and $\Deltat{H}$ is the spectral index of the higher-energy part of $N_{\mu}(E,t)$. % (this symbol has been already used in pancake case in {\S} \ref{sec4_1}, but here we use this in both pancake and isotropic cases)
%  Note that here $\alpha$ is described as a positive value, opposite to the usual definition.

\begin{figure}[htbp]
\centering
\plotone{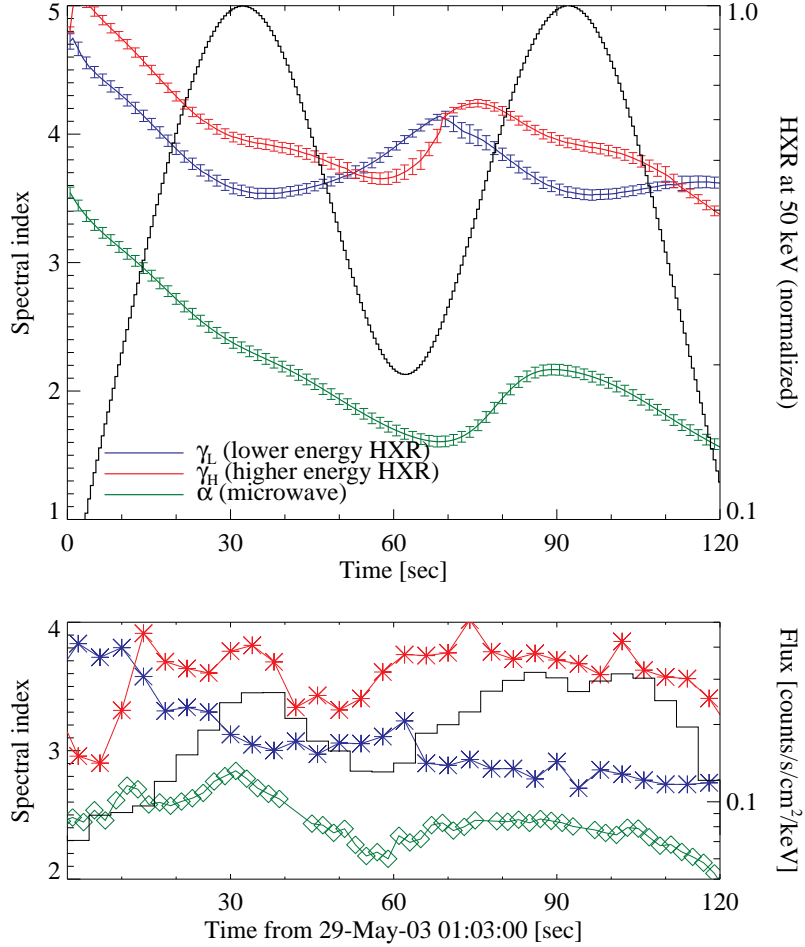}
\caption{Time profiles of the spectral indices of the non-thermal emissions. {\it Upper}: Calculation result in the pancake case. The blue, red, and green lines with error bars are $\gammat{L}{FP}$, $\gammat{H}{FP}$, and $\alphat{FP}$, respectively. The break energy is $\sim 90$ keV. The solid line denotes the predicted HXR flux at 50 keV on a normalized scale. {\it Lower}: Observation result during 01:03 - 01:05 UT. Note that the microwave spectral indices are multiplied by a factor of 2. For a complete set of these data, see Figure \ref{spec}.}
\label{spec_result_pan}
\end{figure}

\begin{figure}[htbp]
\centering
\plotone{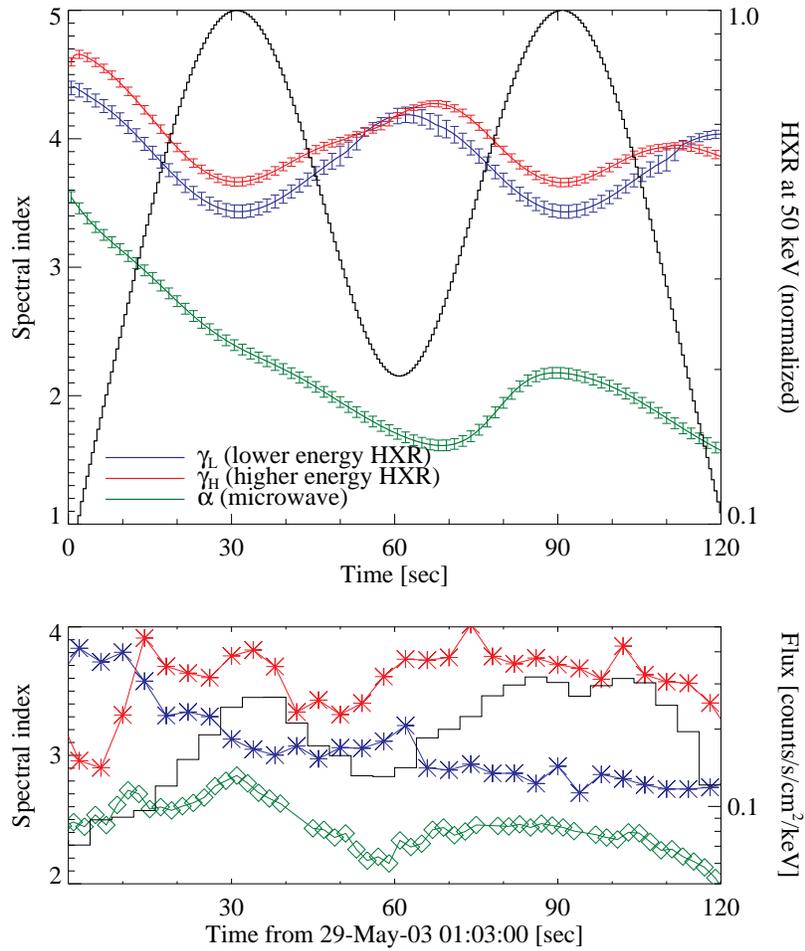}
\caption{Same as Figure \ref{spec_result_pan}, but calculation result is in the isotropic case.}
\label{spec_result_iso}
\end{figure}

Figures \ref{spec_result_pan} and \ref{spec_result_iso} show the time profiles of the spectral indices of the non-thermal emissions. In these figures, the upper plots are the calculation results in the pancake and isotropic cases, respectively. The lower plots are the observation results (but the microwave spectral indices are multiplied by a factor of 2) during 01:03 - 01:05 UT for comparison with the calculations. The complete set of the observation results is in Figure \ref{spec}. Colors (blue, red, and green) denote the lower-energy HXRs, the higher-energy HXRs, and the microwaves, respectively. Hereafter, spectral indices of the lower-energy HXRs, the higher-energy HXRs, and the microwaves in the calculated spectrum are named $\gammat{L}{FP}$, $\gammat{H}{FP}$, and $\alphat{FP}$, respectively.

Our model calculation results show some agreements with the observations. The values of the microwave spectral indices are smaller by $\sim 1.5-2$ than those of the HXR spectral indices in both the pancake and isotropic cases. This result is quantitatively consistent with not only our observations but also with previous reports \citep[e.g.,][]{1997SoPh..175..157S,2000ApJ...545.1116S}. This is understood by using simple analytic formulae that relate the spectral index of the emissions to that of the parent electrons. We assume that the trapped electrons have a power-law energy distribution, $N_{\mu}(E) \propto E^{-\Delta}$. \cite{1985ARA&A..23..169D} gives an empirical relationship of the spectral indices between the {\gyros} emission in the optically thin regime and the parent electron energy distribution at the site,
\begin{eqnarray}
\alpha=0.9 \Delta - 1.22. \label{gs-dulk}
\end{eqnarray}
Although this relationship is derived under the assumption that the parent electrons have an isotropic pitch-angle distribution, we use it for a rough estimate of the microwave spectral index from electrons with an arbitrary pitch-angle distribution. For HXRs, we can use equation (\ref{prec_rate}) that shows the energy distribution of the precipitating electron flux, $F_{\mu}(E) \propto E^{-(\Delta-x)}$.
% Using equation (\ref{prec_rate}), the energy distribution of the precipitating electrons is $F_{\mu}(E) \propto E^{-(\Delta-x)}$.
 An analytic expression for non-relativistic thick-target {\bremss} gives the relationship of the spectral indices between the HXR emission and the parent electron flux precipitating into the thick-target region \citep[e.g.,][]{1972SoPh...24..414H},
\begin{eqnarray}
\gamma = (\Delta-x) -1. \label{thick-express}
\end{eqnarray}
Subtracting equation (\ref{gs-dulk}) from equation (\ref{thick-express}), we find the difference of the spectral indices between the HXR and microwave,
\begin{eqnarray}
\gamma - \alpha = 0.1\Delta -x + 0.22. \label{spec_diff}
\end{eqnarray}
In the weak diffusion regime $x$ takes negative value, around $-1.0$ in our calculation. This eventually yields $\gamma - \alpha \sim 1.5$ (for $\Delta = 3$) which is in agreement with the observations. Thus we conclude that the difference of the spectral indices between the HXR and microwave emissions can be interpreted as a consequence of the parent electron transport in the TPP model in weak diffusion regime.

There is no much difference in the time profiles of $\alphat{FP}$ between the pancake and isotropic cases, in that both show the {\it soft-hard-harder} behavior during the first spike (0-60 sec). Our calculations of the TPP model can successfully reproduce the often observed {\it soft-hard-harder} behavior of the microwave spectrum \citep[e.g.,][]{2000ApJ...545.1116S}. On the other hand, there is a difference in the time profiles of $\gammat{L}{FP}$ and $\gammat{H}{FP}$ between the pancake and isotropic cases. In the isotropic case, the time profiles of $\gammat{H}{FP}$ and $\gammat{L}{FP}$ both show the {\it soft-hard-soft} behavior. In the pancake case, however, $\gammat{H}{FP}$ shows similarity with $\alphat{FP}$ rather than with $\gammat{L}{FP}$, which only shows the {\it soft-hard-soft} behavior. During the decay phase of the earlier spike in the observation (01:03:30 - 01:04:00 UT), both $\gammat{H}{obs}$ and $\alphat{obs}$ show hardening whereas $\gammat{L}{obs}$ shows slight softening. During the rise phase of the later spike in the observation (01:04:00 - 01:04:20 UT), $\gammat{H}{obs}$ and $\alphat{obs}$ show softening. These tendencies can be seen during the respective 30-60 sec and 60-80 sec only in the pancake case calculation (Fig. \ref{spec_result_pan}). These qualitative agreements suggest that it is reasonable to consider the pancake pitch-angle distribution of the injection flux rather than the isotropic one to explain the observed spectral behavior in the 2003 May 29 flare.

\subsection{Validity of our interpretation}\label{sec4_3}
 In our current study, we have utilized the model electron distributions calculated for only two different types of the injection pitch-angle distribution (pancake and isotropic) with nominal values of parameters: the ambient plasma number density $n$ and the loss cone angle cosine $\mu_{\rm c}$ (the mirror ratio). Since these parameters also affect the evolution of electrons, we have to systematically investigate the set of parameters with which we can reproduce the observation of the 2003 May 29 flare. We believe that $n$ of an order of $10^{10}\;{\rm cm^{-3}}$ from the GOES observation and the mirror ratio of $1.2-3$ derived by \cite{1998ApJ...502..468A} are reasonable ranges. A much higher mirror ratio drastically reduces the number of precipitating electrons and the resultant thick-target HXR emissions at the footpoints, which may be in disagreement with the observations. A much higher density $(n \gsim 10^{11}\;{\rm cm^{-3}})$ at the trap region would produce strong coronal thin-target HXR emissions, which is rarely observed. As such, we consider that a relatively lower density at the trap region (less than that of the SXR bright loop) and a relatively smaller mirror ratio should be reasonable, and thus we believe the values adopted in our calculation, $n=3\times 10^{10}\;{\rm cm^{-3}}$ and $\mu_{\rm c}=0.7$, are in reasonable ranges. Our interpretation made earlier in this paper remains valid for the specified ranges of the number density and the mirror ratio. To achieve a better agreement between the calculation and the observation, we need to further refine the model distribution of the injection flux.

\cite{2000ApJ...531.1109L} and \cite{2000ApJ...543..457L} reported the microwave observation of a GOES C2.8 flare on 1993 June 3. They further performed the calculations of the TPP model. They carried out a systematic investigation by varying the number density and injection time scale as well as the injection pitch-angle distribution, to search for the best parameter set that agreed with their observation. They concluded that the electrons are confined to a narrow range ($|\mu| \lsim 0.26$) of pitch-angle and are injected into a low density $(n \sim 4 \times 10^{9}\;{\rm cm^{-3}})$ trap region. This number density is much lower than that we assumed in this paper. This may be because we observed the large X-class flare while they observed the small C-class flare.

Our interpretation of the difference of the spectral indices between the HXRs and microwaves based on the TPP model is subject to the observation that microwaves are emitted at the loop-top whereas HXRs are emitted at the footpoints. Our assumption in the TPP model that the HXR and microwave emitting electrons are treated separately is inadequate for a flare which shows, for example, microwaves as well as HXRs at the footpoints. In such a flare, the microwave emitting electrons are identical to the HXR emitting ones. This means $x=0.5$, yielding $\alpha \simeq \gamma$ in equation (\ref{spec_diff}). The footpoint microwave emissions would thus be expected to have almost the same spectral index as the footpoint HXRs. \cite{2002ApJ...576L..87Y} reported that the footpoint microwave emission has a softer spectral index than the loop-top one by $\sim 2$. Their result is consistent with our interpretation. Simultaneous observations of HXRs and microwaves in a flare showing footpoint HXR and microwave emissions would be useful to further explore the validity of our interpretation.

\section{Conclusion}\label{sec5}
We presented the comparative study of the non-thermal emissions of the flare occurring on 2003 May 29 using the RHESSI HXR and Nobeyama microwave observations. Further, we considered the electron transport model, TPP, to explain the observations.

The 2003 May 29 flare showed two non-thermal HXR sources at the footpoints and a microwave source at the loop-top, as observed with RHESSI and NoRH. We interpreted this in terms of the TPP model. We presented the time profiles of the spectral indices of the higher-energy HXRs $\gammat{H}{obs}$ as well as the lower-energy HXRs $\gammat{L}{obs}$ and microwaves $\alphat{obs}$. The spectra of microwaves and HXRs imply that the microwave emitting electrons have a harder energy distribution than the HXR emitting ones. % The values of the microwave spectral index are smaller by $\sim 2$ than those of the HXR spectral indices, implying that the energy distribution of the microwave emitting electrons is harder than that of the HXR emitting ones.
 We found that the time profile of $\gammat{H}{obs}$ shows similarity with that of $\alphat{obs}$ rather than with $\gammat{L}{obs}$, and is delayed from that of $\gammat{L}{obs}$.

We numerically solved the spatially-homogeneous {\FP} equation for the TPP model to describe the evolution of electrons. Precipitating electrons have a softer energy distribution than the trapped ones in the weak diffusion regime. Differences of the injection pitch-angle distribution especially affect the evolution of the precipitating electrons.

 We calculated the microwave and HXR emissions from the calculated trapped electron distribution and precipitation flux for comparison with the observations. The TPP model in the weak diffusion regime can yield a soft HXR spectrum and a hard microwave spectrum. The calculated difference of the spectral indices between the HXRs and microwaves, $\sim 1.5$, is in agreement with the observations. We further found that a pancake pitch-angle distribution for the injected electrons rather than an isotropic distribution is more adequate to qualitatively explain the temporal variation of $\gammat{H}{obs}$. By comparing the model calculation with the observation, we can constrain the pitch-angle distribution of the injected electrons, which is crucially important for understanding the electron acceleration mechanism in solar flares.

Currently, we are improving our treatment of the TPP model to include the spatial inhomogeneity in the Fokker-Planck equation. Using this, a systematic investigation of the best parameter set to explain the observation is in progress, and will be reported in the future.

\begin{acknowledgments}
We would like to thank T. Terasawa, M. Hoshino, S. Masuda, J. Sato, S. Krucker, H. Hudson, K. Matsuzaki, and T. J. Okamoto for discussions and comments. We also thank K. Shibasaki, M. Shimojo, and A. Asai for discussions and for providing the NoRP and NoRH observation data. We thank A. Caspi for his helpful corrections on our manuscript. We appreciate the anonymous referee's helpful comments and corrections on our manuscript, which much improved the quality of this paper. We acknowledge the whole RHESSI team for providing worthwhile data and sophisticated software. One of the authers (T. M.) is supported by COE program of the University of Tokyo, ``Predictability of the Evolution and Variation of the Multi-scale Earth System: An integrated COE for Observational and Computational Earth Science''.
\end{acknowledgments}

\appendix
\section{Calculation of the gyrosynchrotron emissivity} \label{appen}
A general calculation of the {\gyros} radiation in a magnetized plasma \citep{1969ApJ...158..753R} includes effects such as self-absorption, absorption by ambient plasma, and Razin suppression. These effects significantly contribute at low harmonics ($\nu / \nu_{\rm B} \lsim 10$, where $\nu_{\rm B}$ is the electron gyrofrequency) of the {\gyros} radiation. In our current study, however, only the optically-thin {\gyros} radiation at high harmonics ($10 \lsim \nu / \nu_{\rm B} \lsim 100$) from mildly relativistic electrons ($\Gamma \lsim 10$) is of interest. Under such limited conditions, there is an useful expression given by \cite{1981ApJ...251..727P}. We adopt his formula to predict the microwave emission from the {\FP} calculation results. The approximate expression of the {\gyros} emissivity at a frequency $\nu$ and a viewing angle $\theta$ with respect to the magnetic field, from mildly relativistic electrons with arbitrary energy and pitch-angle distributions, is as follows;
\begin{eqnarray}
j(\nu,\theta,t) = \frac{e^2 \nu_{\rm B}}{c}\left(\frac{\nu}{\nu_{\rm B} \sin^2\theta}\right)
\inty{1}{\infty}d\Gamma \inty{-1}{1}d\mu 
N(\Gamma-1,\mu,t) Y(\theta,\Gamma,\mu) {\cal Z}^{2m}(\theta,\Gamma,\mu), \label{appeq1}
\end{eqnarray}
where $e$ is the elementary charge, and,
\begin{eqnarray}
&& Y = \frac{(\cos\theta-\mu\beta)^2 + (1-z^2)(1-\beta\mu\cos\theta)^2}{(1-z^2)^{1/2}(1-\beta\mu\cos\theta)}, \;\;\; m = \frac{\nu \Gamma}{\nu_{\rm B}}(1-\beta\mu\cos\theta) \nonumber\\
&& {\cal Z} = \frac{z \exp\left[(1-z^2)^{1/2}\right]}{1+(1-z^2)^{1/2}}, \;\;\; z=\frac{\beta\sin\theta (1-\mu^2)^{1/2} }{1-\beta\mu\cos\theta}. \label{appeq2}
\end{eqnarray}

%%%%%%%%%%%%%%%%%%%%%%%%%%%%

%%%%%%%%%%%%%%%%%%%%%%%%%%%%
% \begin{figure}[htbp]
% \centering
% %\includegraphics[clip,scale=0.3,angle=90]{./mapplot.ps}
% \plotone{./spec_sample.ps}
% \caption{RHESSI 40-200 keV count spectrum (points with error bars) and the fitted double power-law model (thick line) during 01:04:28 - 01:04:32 UT in the 2003 May 29 flare. The spectral indices of the lower- and higher-energy part, and the break energy are obtained to be $3.01 \pm 0.001$, $3.85 \pm 0.003$, and $91.8 \pm 0.1 \; {\rm keV}$, respectively.} 
% \label{ex_spectrum}
% \end{figure}
%%%%%%%%%%%%%%%%%%%%%%%%%%%%

%%%%%%%%%%%%%%%%%%%%%%%%%%%%
%%%%%%%%%%%%%%%%%%%%%%%%%%%%

%%%%%%%%%%%%%%%%%%%%%%%%%%%%
% \begin{figure}[htbp]
% \centering
% %\includegraphics[clip,scale=0.3,angle=90]{./specplot.ps}
% \plotone{./specplot.ps}
% \caption{Time profiles of the spectral indices of the non-thermal emissions during the impulsive phase of the 2003 May 29 flare. The blue and red asterisks are the spectral indices of the lower-energy ($< \sim 100$ keV) and the higher-energy ($> \sim 100$ keV) HXR, and the green diamonds are the spectral indices of the microwave in the optically thin part (plotted by an absolute value), respectively. The fitted spectral indices which coefficients of variation exceed 0.1 are plotted by small symbol. The solid line denotes the RHESSI 50-100 keV flux.}
% \label{spec}
% \end{figure}
%%%%%%%%%%%%%%%%%%%%%%%%%%%%

%%%%%%%%%%%%%%%%%%%%%%%%%%%%
%%%%%%%%%%%%%%%%%%%%%%%%%%%%

%%%%%%%%%%%%%%%%%%%%%%%%%%%%
%%%%%%%%%%%%%%%%%%%%%%%%%%%%

%%%%%%%%%%%%%%%%%%%%%%%%%%%%
%%%%%%%%%%%%%%%%%%%%%%%%%%%%

%%%%%%%%%%%%%%%%%%%%%%%%%%%%
%%%%%%%%%%%%%%%%%%%%%%%%%%%%

%%%%%%%%%%%%%%%%%%%%%%%%%%%%
%%%%%%%%%%%%%%%%%%%%%%%%%%%%

%%%%%%%%%%%%%%%%%%%%%%%%%%%%
%%%%%%%%%%%%%%%%%%%%%%%%%%%%

%%%%%%%%%%%%%%%%%%%%%%%%%%%%
% \begin{figure}[htbp]
% \centering
% %\includegraphics[clip,scale=0.3,angle=90]{./cal_ltcplot.ps}
% \plotone{./cal_ltcplot_rev.eps}
% \caption{HXR lightcurves predicted by the {\FP} calculation results in the case of the pancake pitch-angle distribution of the injection flux. The HXR fluxes at 50 keV, 70 keV, 99 keV, 140 keV, and 197 keV in a normalized scale are drawn from top to bottom. The dashed lines denote the peak times of each spike in the HXR flux at 50 keV.}
% \label{cal_ltc}
% \end{figure}
%%%%%%%%%%%%%%%%%%%%%%%%%%%%

%%%%%%%%%%%%%%%%%%%%%%%%%%%%
%%%%%%%%%%%%%%%%%%%%%%%%%%%%

%%%%%%%%%%%%%%%%%%%%%%%%%%%%
% \begin{figure}[htbp]
% \centering
% %\includegraphics[clip,scale=0.3,angle=90]{./emission_specplot.ps}
% \plotone{./emission_specplot2.ps}
% \caption{Time profiles of the spectral indices of the non-thermal emissions predicted by the {\FP} calculation results. The {\it upper} and {\it lower} panels are  the results in the pancake and isotropic case, respectively. The blue and red lines are the spectral indices of the lower-energy ($< \sim 90$ keV) and the higher-energy ($> \sim 90$ keV) HXR, and the green lines are the spectral index of the microwave in the optically thin part (by an absolute value), respectively. The solid lines denote the predicted HXR flux at 50 keV in a normalized scale.}
% \label{emission_spec}
% \end{figure}
%%%%%%%%%%%%%%%%%%%%%%%%%%%%

%%%%%%%%%%%%%%%%%%%%%%%%%%%%
%%%%%%%%%%%%%%%%%%%%%%%%%%%%

%%%%%%%%%%%%%%%%%%%%%%%%%%%%
%%%%%%%%%%%%%%%%%%%%%%%%%%%%

%%%%%%%%%%%%%%%%%%%%%%%%%%%%
% \begin{figure}[htbp]
% \centering
% %\includegraphics[clip,scale=0.3,angle=90]{./prepplot.ps}
% \plotone{./prepplot.ps}
% \caption{The precipitation rate $F_{\mu}$/$N_{\mu}$ at $t=30$ as a function of electron energy. The slope of this line in the higher energy is $\sim$1.4, which is consistent with the dependence of the Coulomb frequency on electron energy.}
% \label{prep}
% \end{figure}
%%%%%%%%%%%%%%%%%%%%%%%%%%%%

\end{document}